\documentclass[acmsmall,screen]{acmart}
\AtBeginDocument{%
  \providecommand\BibTeX{{%
    \normalfont B\kern-0.5em{\scshape i\kern-0.25em b}\kern-0.8em\TeX}}}

\usepackage{graphicx}
\usepackage{xspace}
\usepackage{color}
\usepackage{xcolor}
\usepackage{listings}
\usepackage{amsthm}
\usepackage{amsmath}
\usepackage{subcaption}
\usepackage{amssymb}
\usepackage{stmaryrd}
\usepackage{multirow}
\usepackage{enumitem}
\usepackage{balance} 
\usepackage{natbib}

\usepackage{caption}
\usepackage{booktabs}

\usepackage{adjustbox}

\usepackage{amsthm}
\newtheorem{definition}{Definition}

\usepackage[ruled,vlined,linesnumbered]{algorithm2e}
\setitemize[0]{leftmargin=10pt}

\newcommand{\etal}{\hbox{\emph{et al.}}\xspace}
\newcommand{\eg}{\hbox{\emph{e.g.}}\xspace}
\newcommand{\ie}{\hbox{\emph{i.e.}}\xspace}

\newcommand{\wrt}{\hbox{\emph{w.r.t.}}\xspace}
\newcommand{\etc}{\hbox{\emph{etc}}\xspace}

\newcommand{\tool}{\hbox{\textsc{Genie}}\xspace}
\newcommand{\activeviewinformation}{\hbox{active view information}\xspace}
\newcommand{\totalbugs}{\hbox{34}\xspace}
\newcommand{\changes}[1]{\textcolor{blue}{{#1}}}

\setcopyright{rightsretained}
\acmPrice{}
\acmDOI{10.1145/3485533}
\acmYear{2021}
\copyrightyear{2021}
\acmSubmissionID{oopsla21main-p344-p}
\acmJournal{PACMPL}
\acmVolume{5}
\acmNumber{OOPSLA}
\acmArticle{156}
\acmMonth{10}

\begin{document}

\title{Fully Automated Functional Fuzzing of Android Apps for Detecting Non-crashing Logic Bugs}

\author{Ting Su}
\authornote{This work was mainly conducted when Ting Su was a postdoc at ETH Zurich.}
\author{Yichen Yan}
\affiliation{%
	\institution{East China Normal University}
	\department{Shanghai Key Laboratory of Trustworthy Computing}
	\country{China}
}

\author{Jue Wang}
\affiliation{%
	\institution{Nanjing University}
	\department{State Key Laboratory for Novel Software Technology}
	\country{China}
}

\author{Jingling Sun}
\author{Yiheng Xiong}
\author{Geguang Pu}
\affiliation{%
	\institution{East China Normal University}
	\department{Shanghai Key Laboratory of Trustworthy Computing}
	\country{China}
}

\author{Ke Wang}
\affiliation{%
	\institution{Visa Research}
	\country{USA}
}

\author{Zhendong Su}
\affiliation{%
	\institution{ETH Zurich}
	\department{Department of Computer Science}
	\country{Switzerland}
}

\begin{abstract}

Android apps are GUI-based event-driven software and have become
ubiquitous in recent years. Obviously, functional correctness is critical for an app's success.
However, in addition to crash bugs, \emph{non-crashing functional bugs} (in short as ``non-crashing bugs'' in this work) like inadvertent function failures, silent user data lost and incorrect display information are prevalent, even in popular, well-tested apps.
These non-crashing functional bugs are usually caused by program logic errors and manifest themselves on the graphic user interfaces (GUIs).
In practice, such bugs pose significant challenges in
effectively detecting them because (1) current practices heavily rely on
expensive, small-scale manual validation (\emph{the lack of
	automation}); and (2) modern \emph{fully automated} testing has been limited to
crash bugs (\emph{the lack of test oracles}).

This paper fills this gap by introducing \emph{independent view fuzzing},
\emph{a novel, fully automated approach} for detecting non-crashing functional bugs in Android apps.
Inspired by metamorphic testing, our key insight is to leverage the commonly-held \emph{independent view
property} of Android apps to manufacture property-preserving mutant
tests from a set of seed tests that validate certain app  properties.  
The mutated tests help exercise the tested apps under additional, adverse
conditions. Any property violations indicate likely functional bugs for further manual confirmation.
We have realized our approach as an automated, end-to-end functional
fuzzing tool, \tool.  Given an app, (1) \tool automatically
detects non-crashing bugs without
requiring human-provided tests and oracles (thus
\emph{fully automated}); and (2) the detected non-crashing bugs are
diverse (thus \emph{general and not limited to specific functional
  properties}), which set \tool apart from prior work.

We have evaluated \tool on 12 real-world Android
apps and successfully uncovered \totalbugs previously unknown non-crashing  bugs
in their latest releases --- all have been confirmed, and 22 have
already been fixed.  Most of the detected bugs are nontrivial and have
escaped developer (and user) testing for at least one year and
affected many app releases, thus clearly demonstrating \tool's effectiveness.
According to our analysis, \tool achieves a reasonable true positive rate of 40.9\%, while these \totalbugs non-crashing bugs could not be detected by prior fully automated GUI testing tools (as our evaluation confirms).
Thus, our work complements and enhances
existing manual testing and fully automated testing for crash bugs.



\end{abstract}



\maketitle

\section{Introduction}
\label{sec:intro}

Android apps are GUI-centered event-driven software. The number and
diversity of them have grown rapidly. Recent studies show that app
users highly value user experience --- only 16\% of the users will try
a function-failing app more than twice~\cite{Compuware,localytics}.
Therefore, ensuring functional correctness is critical for improving
an app's success and its user loyalty. However, besides
crashes, non-crashing functional bugs\footnote{Our work focuses on non-crashing functional bugs which are caused by program logic errors and lead to GUI issues. Other non-crashing bugs like performance or energy bugs violate non-functional requirements and thus are not within our scope.} (\eg, inadvertent function failures, silent
user data lost, incorrect display information) are difficult to detect
and frequently escape from standard developer testing. In fact, many
such errors are only noticed by end users post-depolyment, and
sometimes lead to severe consequences in real
life~\cite{pinduoduo,shadow_app}.

Although tremendous progress has been made to improve and automate GUI
testing of Android apps, existing techniques cannot
\emph{fully automatically} detect non-crashing bugs~\cite{RubinovB18,TramontanaAAF19,KongLGLBK19,themis}. For example,
Monkey~\cite{monkey}, Sapienz~\cite{mao_sapienz_2016} and
Stoat~\cite{su_stoat_2017} represent the state-of-the-art GUI testing
techniques of Android
apps~\cite{choudhary_automated_2015,WangLYCZDX18}, but all of them are
limited to crashes, and none can detect non-crashing bugs. Neither
can the state-of-the-practice automatically achieve this. For
example, off-the-shelf test automation frameworks (\eg,
Espresso~\cite{espresso}, Robotium~\cite{robotium}, and
Appium~\cite{appium}) can automate test execution, but require
carefully-designed manual tests (with oracles) to verify functional
correctness. Static analysis tools (\eg, Lint~\cite{lint},
FindBugs~\cite{findbugs}, and Infer~\cite{Infer}) can find generic
coding errors, but are ineffective in detecting app-specific
non-crashing logic bugs.

One \emph{key difficulty} of automatically finding non-crashing bugs is
the lack of test oracles~\cite{BarrHMSY15,VasquezMP17}. Due to
frequent requirement changes, app developers seldom maintain detailed
functional specifications~\cite{how_developers_test_apps}.
As a result, the current practices heavily rely on expensive,
painstaking manual
validation~\cite{kochhar2015understanding,how_developers_test_apps}.
Relevant prior work either leverages the codified oracles in the
developer tests~\cite{FardMM14,AdamsenMM15} or manually defines
oracles for specific app
functionalities~\cite{ZaeemPK14,LamZC17,MarianiPZ18,HuZY18,RosenfeldKZ18,rl_testing}
to detect non-crashing bugs, however with limited usability, effectiveness
and scalability.\footnote{Our investigation of 1,752
Android apps on GitHub reveals that only 62 apps ($\approx 3.5\%$)
contain the GUI-level tests written by developers, and the number of such tests is typically
very small, resulting in limited code coverage.}


\begin{figure*}[t]
	\centering
	\includegraphics[width=1.0\textwidth]{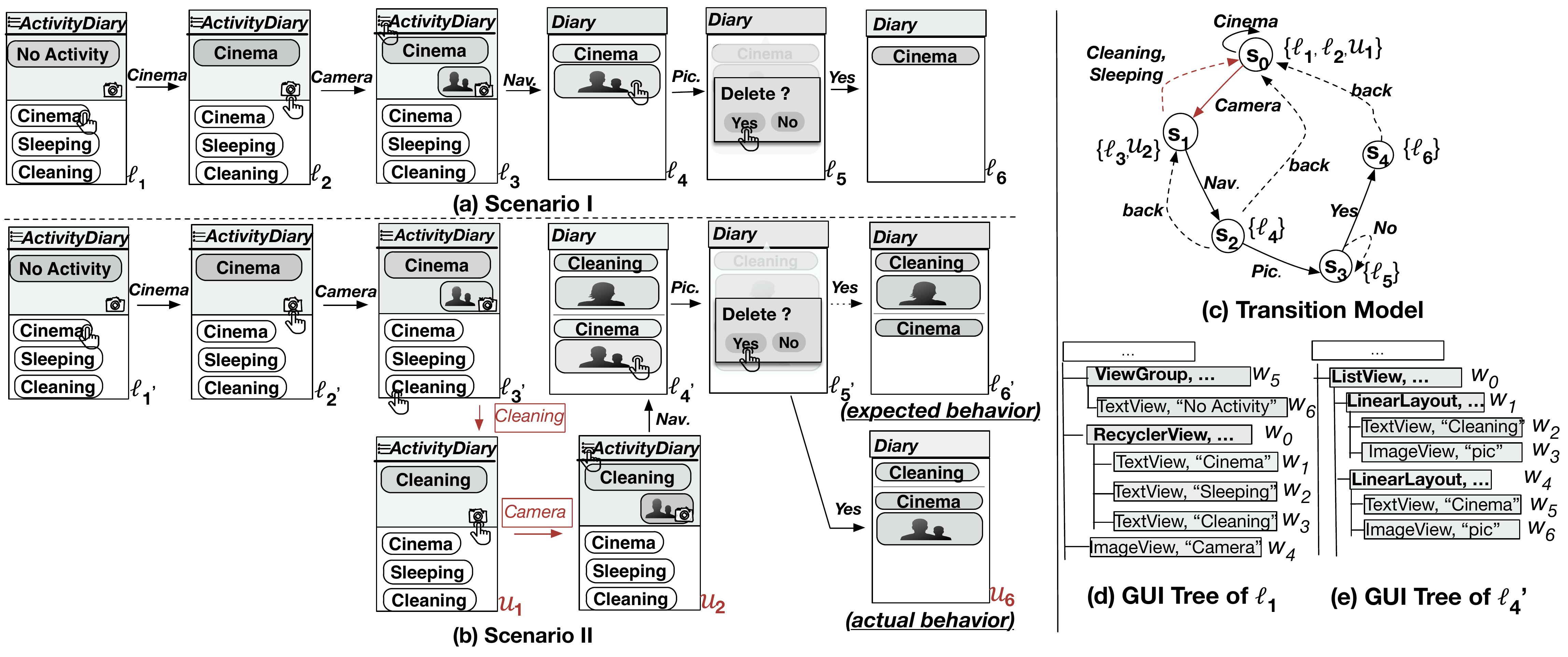}
        \vspace*{-1.8pc}
	\caption{\emph{ActivityDiary}. (a) Scenario I: one typical usage
          scenario. The texts on the arrows denote user events (the
          finger icon is the event location). (b) Scenario II: the expected behavior ($\ell_6'$) \emph{v.s.} actual
          behavior ($u_6$) of \emph{ActivityDiary} by adding 
          two additional events (annotated in the two red boxes) into Scenario I. 
           (c) 
          A partial GUI transitional model. (d) and (e) are the GUI
          trees of $\ell_1$ and $\ell_4$', respectively. }
	\label{fig:activity_diary_example}
\end{figure*} 

\vspace{1pt}
\noindent\textbf{Independent View Property}.
To tackle this challenge, we propose \emph{independent view fuzzing},
\emph{a novel, automated approach} 
to effectively uncovering non-crashing bugs of Android apps without requiring human-provided tests and
oracles, and is not limited to specific functionalities.  Inspired the idea of
metamorphic testing~\cite{chen1998metamorphic,ChenKLPTTZ18,survey_metamorphic_testing}, our key
insight is that many apps hold the \emph{independent view
  property}, that is \emph{interacting with one GUI view (widget) does not
  affect the states of the others and only adds additional
  GUI effects}.  
For example, Fig.~\ref{fig:activity_diary_example}
shows \emph{ActivityDiary}, an app released on Google
Play~\cite{activity_diary}, which is used to record one's daily activities (we have simplified the GUI pages of this app and its workflow).
For example, the three buttons \emph{Cinema}, \emph{Sleeping} and \emph{Cleaning} on $\ell_1$ are some predefined activities for selection, which act like the options in a list 
(these buttons are always visible after selection).
Fig.~\ref{fig:activity_diary_example}(a) shows one of its basic functionality:
a user can start an activity (\eg, \emph{Cinema} in the list on $\ell_1$) and
attach pictures to this activity to record information 
(\eg, the movie the user watched in a cinema on $\ell_3$).
Specifically, the user starts (by clicking) 
\emph{Cinema} ($\ell_1\Rightarrow\ell_2$, ``No Activity'' changes to ``Cinema'') and attaches a picture
to it by clicking the \emph{Camera} button ($\ell_2\Rightarrow\ell_3$, a picture is added). The user can switch to the \emph{diary} page ($\ell_3\Rightarrow\ell_4$), where the app
shows the history of the user's activities in the reverse chronological order (\ie, \emph{Cinema} in this case), 
and the user can delete the picture ($\ell_5\Rightarrow \ell_6$, click ``Yes'').  
Common
knowledge tells us that \emph{Sleeping} or \emph{Cleaning} are obviously 
independent from \emph{Cinema} on page $\ell_3$, and thus
switching to another activity (\eg, \emph{Cleaning}) on page $\ell_3$
following the subsequent events, the deletion of \emph{Cinema}'s
picture will not be affected. 
Fig.~\ref{fig:activity_diary_example}(b)  shows this
expected behavior (denoted by the ending page $\ell_6^{'}$): when the user switches to \emph{Cleaning} ($\ell_3'\Rightarrow u_1$) and takes a picture for it ($u_1\Rightarrow u_2$), \emph{Cinema}'s picture can
still be successfully deleted ($\ell_5^{'}\Rightarrow\ell_6^{'}$).  The only
GUI differences are the additions of \emph{Cleaning} and its picture
on $\ell_4^{'}$ and $\ell_6^{'}$.

Inspired by the above observation, our key idea is to leverage this
independent view property to manufacture app property-preserving
mutant tests from a set of seed tests that witness certain properties
of an app. These mutant tests validate these properties under
additional, adverse conditions.  Any property violations could
indicate likely non-crashing bugs for manual confirmation.  For example,
Fig.~\ref{fig:activity_diary_example}(a) (Scenario I) witnesses one
specific app property, \ie, whether \emph{Cinema}'s picture
can be correctly deleted. Fig.~\ref{fig:activity_diary_example}(b) (Scenario II) can be viewed as a mutant
test that validates this property by inserting two additional
events (\ie, ``click \emph{Cleaning} and take a picture'') at $\ell_3'$.
Intuitively, this app property should hold, otherwise a
likely functional bug happens.  In fact, a real non-crashing 
bug~\cite{activity_diary_bug} did exist. 
Fig.~\ref{fig:activity_diary_example}(b)  shows the actual app
behavior (denoted by the ending page $u_6$), \emph{Cleaning}'s picture was erroneously deleted
while \emph{Cinema}'s picture was still kept. We aim to
identify such critical GUI inconsistencies between the seed test and
its mutant tests to capture likely non-crashing functional bugs.

\vspace{1pt}
\noindent\textbf{Prevalence of the Independent View Property}. 
We find that validating the independent view property can detect many non-crashing bugs.
We selected five popular, open-source apps, namely, \emph{Ankidroid}~\cite{ankidroid}, \emph{AmazeFileManager}~\cite{AmazeFileManager}, \emph{AntennaPod}~\cite{AntennaPod}, \emph{k-9mail}~\cite{k9}, and \emph{TeamNewPipe}~\cite{NewPipe}, and manually inspected all their bugs 
reported within the recent one-year duration. 
Note that these apps are popular (with 1K+ stars on GitHub and 1M+ installations on Google Play), diverse
(covering different app categories), and well-maintained (with many developer-written regression tests and oracles).
It took us over one-person month in this process to manually inspect the bug reports.
We identified 129 non-crashing bugs, 38 of which (29.5\%$\approx$38/129) violated the independent view property (the percentages range from 21$\sim$43\% across these individual apps). 
Thus, the study results clearly show the general applicability of leveraging this property for detecting a large class of non-crashing bugs in real-world apps.
Appendix~\ref{sec:appendix_investigation} details this investigation and the results.


\vspace{1pt}
\noindent\textbf{Independent View Fuzzing}.
To realize the above idea, we face two technical challenges: (1)
\emph{the systematic generation of app property preserving mutants\changes{\footnote{To make it clear, our approach mutates the GUI test inputs rather than the program under test. Throughout the paper, we may use \emph{mutants} for short to denote \emph{mutant tests}.}}}; and
(2) \emph{the precise identification of property violations}.  This paper
proposes two key techniques to overcome these challenges.
The first technique analyzes GUI pages and their layouts during the
execution of a seed test to infer independent views, and leverages a
GUI transitional model to systematically generate independent
events. The second technique compares the GUI effects between the seed
test and each of its mutants to decide property violations. At the
high level, it attempts to capture the intuition that, because of the
inserted events are independent, the GUI effects of the seed test
should be preserved in any mutant test's execution. In other words,
the inserted events should only add, but not remove, GUI effects from
the execution of the seed test.  Otherwise, it indicates a likely
functional bug.



\begin{figure*}[t]
	\centering
	\includegraphics[width=1.0\textwidth]{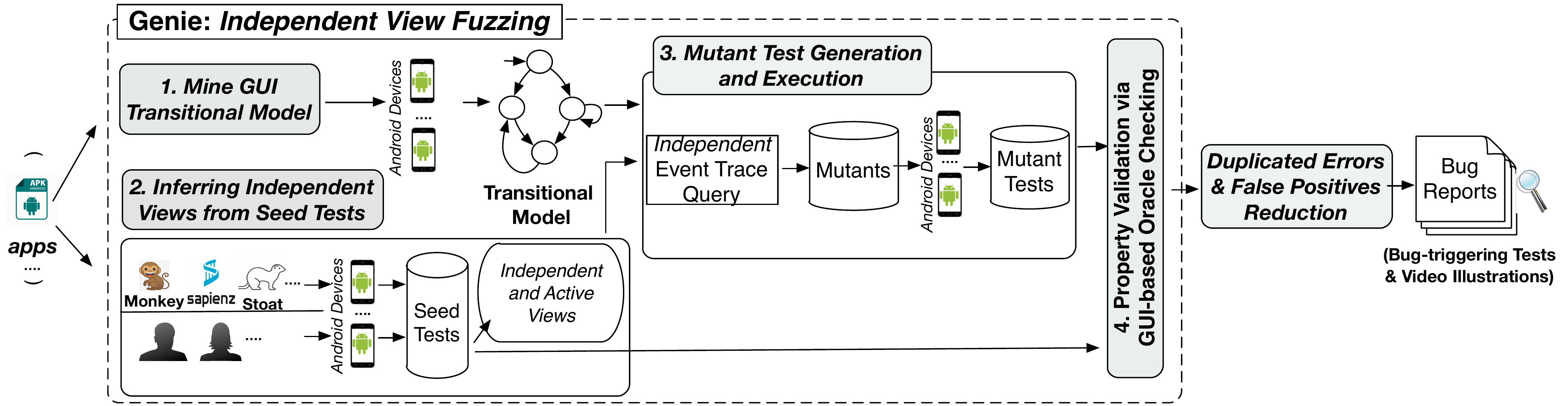}
	\vspace*{-1.0pc}
	\caption{Workflow of our approach.}
	\label{fig:framework_detail}
\end{figure*}

We have realized our techniques as an automated, end-to-end testing tool,
\tool. It can automatically find non-crashing bugs of Android apps
and report them for manual confirmation.
In practice, \tool operates in four main steps as depicted in Fig.~\ref{fig:framework_detail}: 
(1) mining a GUI transitional model from the app,
(2) generating a set of random seed tests and executing each seed test to
infer independent views (\tool could also accept seed tests from
human or existing test generation tools),
(3) leveraging
the independent views and the transitional model to guide the
generation of mutant tests and executing them, and
(4) comparing each
seed test and its corresponding mutant tests to identify property
violations.  
Finally, \tool adopts a bug report reducer to remove duplicated reported errors and trivial false positives, and ranks the remaining distinct bug reports according to their occurrences for manual confirmation.
Section~\ref{sec:example} illustrates how these four main steps work on
a real functional non-crashing bug.
Section~\ref{sec:false_positives} details how the bug report reducer removes duplicated errors and false positives, and ranks the remaining bug reports.

\vspace{1pt}
\noindent\textbf{Results}.
Our evaluation on twelve real-world Android apps
demonstrates the effectiveness of \tool.
To date, \tool has successfully discovered \totalbugs non-crashing bugs, all of
which have been confirmed and 22 have already been fixed.  
All these bugs were discovered in the latest
app releases and previously unknown. Moreover, most of the detected
non-crashing bugs are nontrivial and long latent --- out of the \totalbugs bugs,
26 escaped developer testing for a long duration (1-4 years), while 19
affected more than 10 releases on the app market.  
According to our analysis, \tool achieves a reasonable true positive rate of 40.9\%, while prior fully automated GUI testing tools for Android cannot detect these
\totalbugs non-crashing functional bugs (as our evaluation confirms).

This paper makes the following key contributions:
\begin{itemize}
	\item At the conceptual level, it proposes independent view fuzzing,
		  the first automated approach to detecting non-crashing functional bugs in Android apps
          without requiring human-provided tests and oracles by exploiting the
          commonly-held independent view property of Android apps;
	\item At the technical level, it introduces two novel, effective
          techniques to tackle two challenges: the systematic generation of property-preserving mutants from the provided
          seed tests and the automated GUI-based oracle checking; and
	\item At the empirical level, it presents the realization of
          the approach, \tool, and the evaluation of it on
          twelve real-world apps, leading to \totalbugs previously unknown,
          mostly long latent non-crashing bugs (all confirmed; 22 fixed).
\end{itemize}

\section{Illustrative Example}
\label{sec:example}

We illustrate our approach via the real functional bug of app
\emph{ActivityDiary}, introduced in Section~\ref{sec:intro}.

\vspace*{2pt}
\noindent\textbf{Step 1 (GUI transitional model construction)}.  We
dynamically explore the app's GUI pages and mine a GUI transitional
model to characterize the app's behavior.  In such a model, a node
denotes a GUI state (abstracted from a set of structurally equivalent GUI
runtime layouts), and an edge denotes a GUI event.  For example,
Fig.~\ref{fig:activity_diary_example}(c) shows a partial
transitional model of \emph{ActivityDiary}, which includes the app
behavior of Scenario I (denoted by the solid lines).  In the model,
the runtime GUI layouts $\ell_1$ and $\ell_2$ in Scenario I are mapped
to $s_0$ because of their structural similarity, while $\ell_3$ is
mapped to a different state $s_1$ because it has an additional picture
view.  Section~\ref{sec:model} defines the state abstraction criterion
and explains in detail how to mine such transitional models.


\vspace*{2pt}
\noindent\textbf{Step 2 (Independent view inference)}.  We generate a
set of random seed tests as the basis for property validation. During
the execution of a seed test, we examine each page's GUI layout (\ie,
a GUI tree) to infer independent views within each group
view. Specifically, a \emph{group view} is a type of GUI views (\eg,
\texttt{ListView} and \texttt{RecyclerView}) for arranging/displaying
a number of \emph{relevant, but functionally independent} child views,
suggested by GUI design guidelines~\cite{material_design}.  On each
layout, we also maintain the \emph{\activeviewinformation} within
each group view to track which view is currently active (\ie,
executed) and update this information across subsequent layouts when
possible. This enables us to infer independent events at any layout of
the seed test with the previous execution information.

For example, we take Scenario I in
Fig.~\ref{fig:activity_diary_example}(a) as a seed test, and execute
it.  On the layout $\ell_1$, view \emph{Cinema}, \emph{Sleeping}, and
\emph{Cleaning} are independent views.  Because from
Fig.~\ref{fig:activity_diary_example}(d), $\ell_1$'s simplified GUI
tree, we observe that \emph{Cinema}, \emph{Sleeping}, and
\emph{Cleaning} are three independent siblings rooted by the same
group view \texttt{RecyclerView}.  Specifically, on layout $\ell_1$,
\emph{Cinema} will be annotated as \emph{active} within its group view
when it is clicked.  The other two views, \emph{Sleeping} and
\emph{Cleaning}, are by default annotated as \emph{inactive}.  This
information will also be updated on $\ell_2$ and $\ell_3$, but will
not be updated on $\ell_4$ (because \emph{Cinema} does not exist on
$\ell_4$). On $\ell_4$, \emph{picture} will be annotated as active
when it is clicked. Note that on $\ell_2$, \emph{Camera} will not be
annotated as active when clicked because it does not belong to any
group view (see Fig.~\ref{fig:activity_diary_example}(d)). In this
way, we can know, on each layout, which views are active or inactive
within each group view.  Section~\ref{sec:seed_test} details how
we infer independent views and maintain their \activeviewinformation
on each GUI layout during the execution of seed tests. 

\vspace*{2pt}
\noindent{\textbf{Step 3 (Mutant test generation and execution)}.  In
	this step, we leverage the independent view property to
	systematically generate independent event traces for creating mutant
	tests.  Specifically, we (1) pick a \emph{pivot layout} from the
	seed test, and use the information from Step~2 to select an
	\emph{inactive} view from any group view as the start of an
	independent event trace, and (2) use the transitional model to guide
	the generation of remaining events and return back to this pivot
	layout. Conceptually, we assume that such a ``loop trace'' will
	exercise some independent functionalities \wrt the subsequent events
	in the seed test, and thus preserve the app behavior of
	the seed test.
	
	For example, if we pick $\ell_3$ (in
	Fig.~\ref{fig:activity_diary_example}(a)) as the pivot layout, based
	on the \activeviewinformation, we know that \emph{Cleaning} is
	inactive and independent from \emph{Cinema}.  Then, we start from
	\emph{Cleaning}, and query the transitional model (see
	Fig.~\ref{fig:activity_diary_example}(c)) to obtain any possible
	event traces that can return back to layout $\ell_3$.  Here, one
	possible independent event trace is ``click \emph{Cleaning} and take a
	picture'' (denoted by the red lines).  In this way, we can
	successfully create the mutant test, \ie, the test of Scenario II in
	Fig.~\ref{fig:activity_diary_example}(b), which can manifest the
	bug.  Section~\ref{sec:mutant_test} details how we systematically
	generate mutant tests by leveraging the \activeviewinformation and
	the transitional model.

\vspace*{2pt}
\noindent{\textbf{Step 4 (GUI-based oracle checking)}.  This step
	computes the differences of GUI effects between the seed test and
	each of its mutants to identify possible property violations, thus
	functional bugs.  Specifically, we use the differences of GUI views
	(denoted by $\Delta$) between two GUI pages to characterize the GUI
	effects of executing the event trace in between.  This formulation
	is general and enables us to perform flexible oracle checking.  If
	$\Delta$ of the seed test is not contained in $\Delta'$ of the
	mutant test (\ie, $\Delta \nsubset \Delta'$), a likely functional
	bug is detected.
	
	For example, in Fig.~\ref{fig:activity_diary_example}(a), by
	comparing layouts $\ell_4$ and $\ell_6$, the GUI effect of the event
	trace between $\ell_4$ and $\ell_6$ can be characterized by the
	deletion of a picture under \emph{Cinema} (denoted as
	$\Delta$).  Similarly, in Fig.~\ref{fig:activity_diary_example}(b),
	we can know the GUI effect of the event trace between $\ell_4'$ and
	$u_6$ can be characterized by the deletion of a picture under
	\emph{Cleaning} (denoted as $\Delta'$).  Obviously, $\Delta
	\nsubset \Delta'$, \ie, these two GUI effects are not the same --- the
	picture was deleted from different activities.  Thus, we detect a
	likely functional bug and report it for manual confirmation, which is indeed
	a real functional bug.  Section~\ref{sec:oracle} details how we
	formulate and perform this GUI-based oracle checking.

\section{Independent View Fuzzing}

This section presents the formulation and technical details of our
approach. It starts by introducing the needed notations and
definitions, and then describes the approach's four core steps.

\subsection{Notations and Definitions}
\label{sec:definition}
Given an Android app, our goal is to generate property-preserving
mutant tests from a set of seed tests. 
An Android app is a 
GUI-centered event-driven program $P$. Each of its GUI pages
is a runtime GUI layout $\ell$, \ie, a GUI tree $T$.
Each node of this tree is a GUI view (or widget) $w$ ($w \in \ell$).
Specifically, each view $w$ has a type $w.type$, which denotes
its view type. For example, a leaf view's $w.type$ can be a 
button (\texttt{Button}) or a text field (\texttt{EditText}).
A non-leaf view's $w.type$ can be a layout view, \eg, 
\texttt{ListView}, \texttt{LinearLayout}, which arranges or displays
other views in specific orders and appearances.
When $P$ is not in the foreground, we define $\ell = \bot$.

A GUI event $e=\langle t, r, o \rangle$ is a tuple, where $e.t$
denotes $e$'s event type (\eg, \texttt{click}, \texttt{edit},
\texttt{swipe}, or a \texttt{system} event),
$e.r$ is a function to obtain $e$'s receiver view (specifically, 
we use $e.r(\ell)= w$ to denote that $e$ can be executed on 
the view $w$ $\in$ $\ell$ at runtime), and $e.o$ denotes the optional data
associated with $e$ (\eg, a string/number for \texttt{edit}). 

A GUI test is a sequence of events $E=[e_1, e_2, \ldots, e_n]$.
$E$ can be executed on $P$ to obtain an execution trace 
$\Pi_{P}(E)=\langle L, I, H \rangle$, where $L$ is a sequence
of runtime layouts $L=[\ell_1, \ell_2, \ldots, \ell_{n+1}]$,
$I$ and $H$ (a map) record the independent views and the active view information
for each layout, respectively (discussed in
Section~\ref{sec:seed_test}). Specifically, we use 
$\ell_{i+1} = e_{i}(\ell_{i})$ to denote the execution of $e_i$,
where $\ell_{i+1}$ is the layout due to the execution of
$e_{i}$ on $\ell_{i}$. In practice, we use these runtime
GUI layouts to approximate app states~\cite{YuanM07}.

Next, we describe the details of our approach's four steps as
illustrated in Section~\ref{sec:example}.

\subsection{Mining GUI Transitional Model}
\label{sec:model}

This step aims to mine a GUI transitional model to represent an app's behavior. 
This model is represented as a tuple $M=\langle S, \Sigma,
\delta\rangle$, where $S$ denotes a set of abstract states $\{s|$
$s$ is a subset of layouts in $L \}$, $\Sigma$ the set of events that
$P$ accepts, and $\delta: S \times \Sigma \rightarrow S$ the transition
function with each transition $\langle s, e, s'\rangle \in \delta$
denoting a state transition from $s$ to $s'$ triggered by $e$.

In practice, to balance the model size and its
precision, we designed a variant of the C-Lv4 GUI comparison
criterion of AMOLA~\cite{BaekB16} to decide whether two runtime layouts $\ell$ and
$\ell'$ are structurally equivalent. If $\ell$ and $\ell'$ are
structurally equivalent, they are grouped together as one state $s$ in
$M$ to mitigate state explosion.  
This criterion compares view
composition between two layouts to decide equality. 
We define the \texttt{Equivalent-with} relation (denoted by
$\simeq$) below.

\begin{definition}
\noindent\textbf{\texttt{Equivalent-with} ($\simeq$)}. Let
layout $\ell$ be represented as $\{w_1, w_2, \ldots, w_{|l|}\}$ 
($|\ell|$ is the number of widgets in $\ell$),
we abstract $\ell$ into $\ell_{abs} = \bigcup_{w \in \ell} \{ w.type \}$, 
which approximates
the structure information of $\ell$ via a type abstraction that
differentiates views through their types.
\footnote{Our implementation
associates each view's type $w.type$ with its
\texttt{resource-id} and \texttt{content-description} from the
corresponding GUI tree node, but omits \texttt{text} and other
attribute values (\eg, \texttt{Clickable}).}
Given layout $\ell$ and $\ell'$, if $\ell_{abs} = \ell'_{abs}$,
\texttt{Equivalent-with($\ell$,$\ell'$)} holds, \ie, $\ell
\simeq \ell'$.
\end{definition}

\noindent\emph{\underline{Example}}. In Fig.~\ref{fig:activity_diary_example}(a), 
$\ell_1 \simeq \ell_2$ because their structures are equivalent (both are similar to Fig.~\ref{fig:activity_diary_example}(d)).
$\ell_2 \not\simeq \ell_3$ because $\ell_3$ has a
picture view.
For $\ell_4$ in Fig.~\ref{fig:activity_diary_example}(a) and $\ell_4'$ in Fig.~\ref{fig:activity_diary_example}(b), 
$\ell_4 \simeq \ell_4'$  
because a \texttt{ListView} will have only two abstract states, 
\ie, empty and non-empty (with any number of children) under this relation.
Fig.~\ref{fig:activity_diary_example}(e) shows the structure of $\ell_4'$.
Fig.~\ref{fig:activity_diary_example}(c) shows the GUI model $M$.

\vspace{1pt}
\noindent\textbf{State coverage-optimized model mining}.  
We propose a \emph{state
coverage-optimized} GUI mining algorithm for constructing the transitional model.
The basic mining process is, 
from a given GUI layout $\ell$, we select and execute an
event $e$ on its receiver view $e.r(\ell)$.  Then, $e$ takes us to a
new layout $\ell'$, from which we continue to select the next
event. During this process, a self-loop transition $(s, e, s')$ is
added into $M$ if $\ell_{abs} \simeq \ell_{abs}'$
(\ie, \texttt{Equivalent-with($\ell$,$\ell'$)}). 
Otherwise, a transition
connecting the two (different) states is added.
Here, $s$ and $s'$ are the states in $M$ corresponding
to the abstract layouts $\ell_{abs}$ and $\ell_{abs}'$, respectively.
If $\ell' = \bot$,
we restore $P$ to the foreground and continue. To improve
the mining performance, the model mining algorithm synergistically combines two existing strategies, \ie, \emph{systematic event selection} (adapted and significantly extended from Stoat~\cite{su_stoat_2017}'s
weighted GUI exploration strategy) and \emph{random event selection} (mimicking Android Monkey~\cite{monkey}'s random exploration strategy).
\begin{itemize}
	\item \emph{Systematic event selection}. 
	This strategy prioritizes an event $e$ if $e$ is
	less frequently executed and can transit to a new layout
	with more new events. 
	Specifically, we assign each event
	$e$ a weight $\Upsilon(e)$ to capture these two pieces of
	information. Assume we are at the $i$th round of event execution,
	\ie, $\ell_{i+1} = e_{i}(\ell_{i})$, we first recursively
	update the weight $\Upsilon(\cdot)$ for each event $e$
	in the worklist by 
	Formula~(\ref{eq:action_weight}) (the worklist stores all the events
	that are identified on previous runtime GUI layouts before
	$\ell_{i+1}$, \ie, $[\ell_1, \ldots, \ell_{i}]$), and then select the event
	with the largest weight on layout $\ell_{i+1}$:
	
	\begin{equation}\label{eq:action_weight}
		\Upsilon_{i+1}(e)=\frac{\Upsilon_{i}(e)+ \sum \Upsilon_{i}(new\_events(e))}{exec\_times(e)^2}
	\end{equation}
	where, $new\_events(\cdot)$ denotes the set of events on the
	layout due to the execution of $e$ in the history, $exec\_times(\cdot)$
	records the executed times of $e$ during mining ($i \geq 1$). Initially, $\forall e \in E. \Upsilon_1(e)=100, exec\_times(e)=1$, and $new\_events(e)=\emptyset$. 
	
	\item \emph{Random event selection}. This strategy randomly selects
	events according to their types with
	predefined probabilities (our implementation sets 60\% for \texttt{touch} events, 35\% for \texttt{long-touch} events, and 5\% for \texttt{navigation} events). It reduces the possibility
	of exploration being trapped on a specific GUI page 
	(\eg, repeatedly selecting events from a long \texttt{ListView}
	with similar items).
	
	
\end{itemize}

During model mining, we interleave these two strategies to combine
their effectiveness like hybrid symbolic execution~\cite{MajumdarS07}.
Indeed, we first apply systematic event selection and attempt to cover
as many app states as possible. Upon saturation after a certain number
of rounds, we apply the random strategy to discover possible new
states, and then switch back to systematic event selection.

\noindent\emph{\underline{Example}}. Fig.~\ref{fig:activity_diary_example}(c)
shows a partial transitional model of \emph{ActivityDiary}.
The nodes denote the abstract states,
while the solid edges are the events
from Scenario I (Fig.~\ref{fig:activity_diary_example}(a)) and
the dotted edges are the events from other explorations.

\subsection{Inferring Independent Views from Seed Tests}
\label{sec:seed_test}

\newcommand{\indep}{\perp \!\!\! \perp}

Given a seed test, this step aims to infer independent views
on each layout and maintain their active view information along the
execution trace.  This information will be leveraged to manufacture
property-preserving mutant tests in the next step
(Section~\ref{sec:mutant_test}).  The key of this analysis is, given a
GUI layout, to identify which views are independent from each other
and which views are active according to previous event executions.
To this end, we define the \texttt{Independent\_from} relation
(denoted by $\indep$) and the concept of \emph{active views}.

\begin{definition}
\noindent\textbf{\texttt{Independent\_from} ($\indep$)}.  Conceptually, a GUI
layout $\ell$ can be partitioned into non-overlapping regions that are
rooted by a set of group views $G=\{w_{g_1}, w_{g_2}, \ldots,
w_{g_n}\}$, where $w_{g_i} \in \ell$ is a non-leaf layout view of
\emph{group view} type.\footnote{According to GUI design
	guidelines~\cite{material_design}, the typical group view types
	include \texttt{RecyclerView}, \texttt{ListView}, \texttt{GridView},
	\texttt{ViewGroup}, \texttt{RadioGroup}, \texttt{LinearLayout},
	\texttt{GridLayout}, \etc.}  These regions serve different
functional purposes.  \texttt{Group}($w$) denotes the group
view of $w$. We define view $w_1 \in \ell$ and $w_2 \in
\ell$ to be independent, \ie, $w_1 \indep w_2$, if

\begin{enumerate}[wide, labelwidth=!, labelindent=0pt]
	\item $w_1$ and $w_2$ are the children of two different
	group views in $G$, \ie, $\texttt{Group}(w_i) \neq \texttt{Group}(w_j)$.
	
	\item $w_1$ and $w_2$ are the children of the same group view,
	\ie, $\texttt{Group}(w_i)=\texttt{Group}(w_j)$, and satisfy that $w_1$ and $w_2$
	are siblings\footnote{To ease explanation, we assume $w_1$ and
		$w_2$ are leaf views. When $w_1$ and $w_2$ are non-leaf views,
		their respective children are independent from each other.
		Our implementation considers both cases.} and $w_1.type=w_2.type$.
\end{enumerate}
Note the views that are not within any group view are assumed
independent from those views in group views. We say any views
that are independent from others are \emph{independent views}.
\end{definition}

\noindent\emph{\underline{Example}}. 
Fig.~\ref{fig:activity_diary_example}(d) shows the GUI tree of
$\ell_1$, where $w_1$ (\emph{Cinema}), $w_2$ (\emph{Sleeping})
and $w_3$ (\emph{Cleaning}) are independent views and rooted
by the group view $w_0$ (\texttt{RecyclerView}). Obviously,
clicking one of them will not affect the states of others. 
$w_6$ (\emph{No Activity}) $\indep$ $w_1$ 
holds because $w_6$ is in another group view $w_5$ (\texttt{ViewGroup}).
$w_4$ (\emph{Camera}) $\indep$ $w_1$ also holds because
$w_4$ is not within a group view.
Fig.~\ref{fig:activity_diary_example}(e) shows the GUI tree
of $\ell'_4$, where $w_1$ (\ie, \emph{Cleaning} and its picture)
$\indep$ $w_4$ (\emph{Cinema} and its picture) holds
because they are rooted by the group view $w_0$ (\texttt{ListView}). 
Obviously, clicking \emph{Cinema}'s picture will not affect the
state of \emph{Cleaning}'s picture.

\begin{definition}
\noindent\textbf{Active View}. 
Let $G=\{w_{g_1}, w_{g_2}, \ldots, w_{g_n}\}$ be the set
of group views of layout $\ell$ and $h_\ell \in H $ be the map that
maintains the active view information of $\ell$,
we define the \emph{active view}
within each group view as follows:
\begin{enumerate}[wide, labelwidth=!, labelindent=0pt]
	\item A view $w_i$ becomes \emph{active} when
	an event $e$, satisfying $e.r(l)=w_i$, was executed on $\ell$.
	The \activeviewinformation of $\ell$ is updated as
	$h_\ell[\texttt{Group}(w_i)] = w_i$.
	
	\item A view $w_i$ becomes \emph{inactive}
	when an event $e$, satisfying $e.r(l)=w_j \land \texttt{Group}(w_i)
	= \texttt{Group}(w_j)$,
	was executed on $\ell$. The active view of
	$\ell$ is updated as $h_\ell[\texttt{Group}(w_j)] = w_j$.
	
	\item A view $w_i$ stays \emph{active} or
	\emph{inactive}, when an event $e$, satisfying $e.r(l)=w_j
	\land \texttt{Group}(w_i) \neq \texttt{Group}(w_j)$, was executed on $\ell$.
	The active view of $\ell$ is updated as $h_\ell[\texttt{Group}(w_j)] = w_j$.
	
\end{enumerate}
\end{definition}
Note that those views not within any group view are always assumed \emph{inactive}.  
More importantly, we maintain the
active view information along the execution trace, \ie, update this
information of $\ell$ to one subsequent layout $\ell'$ if similar
views can be located on $\ell'$. This enables us to leverage the
independent view property at any layout of a seed test informed by
previous event executions.

\vspace{2pt}
\noindent\emph{\underline{Example}}. 
Assume Scenario II in Fig.~\ref{fig:activity_diary_example}(b) denotes a seed test.
\emph{Cinema} will become \emph{active}
when it is clicked on layout $\ell_1'$. \emph{Sleeping} and
\emph{Cleaning} are by default \emph{inactive}.
\emph{Cleaning} will become \emph{active} when it is clicked on \emph{$\ell_3'$},
while \emph{Cinema} will become \emph{inactive}.
Because \emph{Cinema} and \emph{Cleaning} are in the same group view.
During event executions, this \activeviewinformation will be updated
from $\ell_1'$ to $\ell_2'$
(\ie, \emph{Cinema} is active on $\ell_2'$), and from $\ell_3'$
to $u_1$ and $u_2$ (\ie, \emph{Cleaning} is active on $u_1$ and $u_2$) since
we can locate the similar view \emph{Cinema} on $\ell_2'$ and
\emph{Cleaning} on $u_1$ and $u_2$, respectively.
But it will not be updated on $\ell_4'$ because \emph{Cleaning}
does not exist on $\ell_4'$. 
On the other hand, clicking \emph{Camera} on $\ell_2'$ and $u_1$
will not affect the \activeviewinformation of \emph{Cinema} or \emph{Cleaning}
as they are in different group views. 
\emph{Camera} itself is always \emph{inactive} because it is not within
any group views.

\vspace{2pt}
To maintain the \activeviewinformation, we define the \texttt{Similar\_with}
relation (denoted by $\sim$) for locating similar views between two GUI layouts.
It is also used for mutant execution in Section~\ref{sec:mutant_test}.

\begin{definition}
\noindent\textbf{\texttt{Similar\_with} ($\sim$)}.  Let $\ell = \{w_1,
\ldots, w_{|\ell|}\}$ be the layout on which event $e$ is
executed, and $\ell' = \{w_1', \ldots, w_{|\ell'|}'\}$ be
another layout. Obviously, there exists a receiver
view $w_i \in \ell$ satisfying $e.r(\ell)=w_i$.  If there exists a
``similar'' view $w_j' \in \ell'$ such that $w_j'$ should be the
receiver view of $e$ on $\ell'$, \ie, $e.r(\ell')=w_j'$, we say
\texttt{Similar\_with($w_i$,$w_j'$)}, \ie, $w_i \sim w_j'$.
\end{definition}

\noindent{\textbf{Locating similar views}}.
Specifically, to locate the similar view $w_j' \in \ell'$ \wrt
$w_i \in \ell$, we first convert the type information of $w_i$
(including its \texttt{text} if it exists) into a string. Then,
we use string equality to check whether there exists at least
one exactly matched view on $\ell'$. If no such views exist
on $\ell'$, we conclude no similar views can be found.
Otherwise, we use the type information of $w_i$ (including
all its children views) to find the ``most similar'' 
view $w_j' \in \ell'$ in terms of string edit distance.

Specifically, we update the \activeviewinformation for an
independent view from $\ell$ to $\ell'$ only if we can locate
a similar view, otherwise we abandon that update.

\vspace{2pt}
\noindent{\textbf{Inferring independent views from seed tests}}.
Algorithm~\ref{algo:seed_test_execution} gives the process
of inferring independent views and maintaining their \activeviewinformation
during the execution of a seed test $E$.
The seed test was randomly generated in our approach (detailed in Section~\ref{sec:seed_generation}). 
Recall that, a seed test $E$'s execution trace is $\Pi_{P}(E)=\langle L, I, H \rangle$,
where $L$ is the sequence of layouts, $I$ and $H$ record the
independent views and the \activeviewinformation for each layout,
respectively. The algorithm annotates the receiver view of
$e$ as active (if it is within a group view) on the current
layout (Line 4). Then, it sends event $e$ to the app, which
transitions to a new layout (Lines 5-6). Next, it updates the
\activeviewinformation from prior GUI layouts to the new
layout (Line 7). The function \texttt{Update\_active\_views}
internally updates the \activeviewinformation from the most
recent similar GUI layout to the current one (Lines 11-16). 
\texttt{Similar\_layout\_type} (Line 12) checks whether two layouts
are of similar type (\ie, denoting similar functional page),
which is a coarse version of \texttt{Equivalent\_with}. 
In practice, \texttt{Similar\_layout\_type} leverages the layout
information to characterize the page type. For example, in Fig.~\ref{fig:activity_diary_example}(a), $\ell_2$ and $\ell_3$
have the similar layout type
because they denote the same \emph{main} page (although they are 
not structurally equivalent). 
Thus, we can update the active information
from $\ell_2$ to $\ell_3$. 
The active view information is updated according to the \texttt{Similar\_with} relation (Lines 13-15).

\begin{algorithm}[t]
	\small
	\newcommand\mycommfont[1]{\small\sffamily\textcolor{gray}{#1}}
	\SetCommentSty{mycommfont}
	\DontPrintSemicolon
	\SetKwFunction{execution}{Execute$_P$}
	\SetKwFunction{getgui}{GetGUI}
	\SetKwFunction{getIndependentView}{Get\_independent\_views}
	\SetKwFunction{sendevent}{SendEventToApp}
	\SetKwFunction{similarType}{Similar\_layout\_type}
	\SetKwFunction{similarwith}{Similar\_with}
	\SetKwFunction{update}{Update\_active\_views}
	\SetKwFunction{reverse}{reverse}
	\SetKwFunction{group}{Group}
	\SetKwProg{func}{Function}{:}{}
	
	\func{\execution{Seed test $E$}}{
		$\ell \leftarrow \getgui() $; $L \leftarrow [\ell]$;  $I \leftarrow []$; 
		$I \leftarrow I::\getIndependentView(\ell)$;
		$H \leftarrow []$ \;
		\ForEach{$e \in E$}{
			
			$ h[\group(e.r(l))] \leftarrow e.r(l) $; $H \leftarrow H :: [h]$ \;
			\sendevent($e.t$, $e.r(\ell)$, $e.o$)\;
			$\ell \leftarrow \getgui()$\; 
			$h \leftarrow \update(\ell, \langle L, I, H \rangle)$\;
			$L \leftarrow L :: [\ell]$; $I \leftarrow I::\getIndependentView(\ell)$; 
		}
		\Return $\langle L, I, H \rangle$\;
	}
	
	\func{\update{layout $\ell$, trace $\langle L, I, H \rangle$}}{
		\tcp*[h]{update the active view info from the prior layouts}\;
		\ForEach{$\langle \ell_i, h_i \rangle \in \langle L, H \rangle.\reverse()$}{
			\If{\similarType($\ell_i$, $\ell$)}{
				\tcp*[h]{Group($w_a$)=$w_g$, and $w_a$ is active}\;
				\ForEach{$\langle w_g, w_a\rangle \in h_i$}{
					\If{$(\exists w_g' \in \ell$.$w_g \sim w_g'$) $\land$ $ (\exists w_a' \in \ell$.$w_a\sim w_a')$}{
						\tcp*[h]{update the active view info}\;
						$h[w_g']$ $\leftarrow$ $w_a'$
					}
				}
				\Return $h$
			}
		}
	}
	\caption{Maintaining independent views and their active view information during the execution of a seed test.}
	\label{algo:seed_test_execution}
\end{algorithm}

\subsection{Mutant Test Generation and Execution}
\label{sec:mutant_test}
This step inserts independent event traces at a given pivot
layout of a seed test to manufacture mutant tests. 
Formally, let $E=[e_1, e_2, \ldots, e_n]$ be a seed test, and
$L=[\ell_1, \ell_2, \ldots, \ell_{n+1}]$ be the 
runtime layouts yielded by $E$, we can create a mutant
test $E'=[e_1, \ldots, e_{i-1}, e_1', e_2', \ldots, e_m', e_i, \ldots, e_n]$ by
inserting an event trace $\tau=[e_1', e_2', \ldots, e_m']$
at a given pivot layout $\ell_{i} \in L$ ($\ell_i = e_{i-1}(\ell_{i-1})$).
Importantly, $\tau$ should satisfy
(1) independent from subsequent events; and (2)
connecting with the event traces before and after the insertion position.

\begin{algorithm}[t]
	\small
	\newcommand\mycommfont[1]{\small\sffamily\textcolor{gray}{#1}}
	\SetCommentSty{mycommfont}
	\DontPrintSemicolon
	\SetKwFunction{generate}{Generate\_mutants}
	\SetKwFunction{getIndependentView}{Get\_independent\_views}
	\SetKwFunction{execution}{Execute$_P$}
	\SetKwFunction{getstate}{Find\_equivalent\_state}
	\SetKwFunction{gentrace}{Search\_traces}
	\SetKwFunction{update}{Update\_active\_views}
	\SetKwFunction{randompop}{Pop}
	\SetKwFunction{push}{Push}
	\SetKwFunction{inactive}{Is\_inactive}
	\SetKwFunction{assemble}{Assemble}
	\SetKwFunction{lastlayout}{Get\_last\_layout}
	\SetKwFunction{maxnum}{MAX\_NUM}
	\SetKwFunction{maxlen}{MAX\_LENGTH}
	\SetKwFunction{reverse}{reverse}
	\SetKwFunction{tail}{tail}
	\SetKwProg{func}{Function}{:}{}
	$\langle L, I, H \rangle \leftarrow \execution(E)$; $mutants \leftarrow []$\; 
	\func{\generate{Seed test $E$, Model $M$}}{
		
		\ForEach{$\langle \ell_i, h_i \rangle \in \langle L, H \rangle$}{
			$\ell_{abs} \leftarrow \getstate(M, \ell)$\; 
			$independent\_traces \leftarrow \gentrace(i, \ell_{abs}, M)$\;
			$tests \leftarrow \assemble(E, \ell, independent\_traces)$\;
			$mutants \leftarrow mutants :: tests$
		}
		\Return $mutants$\;
	}

	\func{\gentrace{insert position $i$, abstract layout $\ell_{start}$, Model $M$}}{
		$ candidate\_traces \leftarrow [] $; $ stack \leftarrow \{\} $;
		
		\ForEach{$\langle \ell_{start}, e, \ell_{next} \rangle \in M[\ell_{start}]$}{
			\tcp*[h]{Get the transitions starting from $\ell_{start} \in M$}\;
			\If{$e.r(L[i]) \in \getIndependentView(L[i]) \land \inactive(e)$}{
				\tcp*[h]{Pick $\tau$'s the starting event from inactive views on $\ell_{i}, \ie, L[i]$}\;
				
				$ stack.\push\{[e]\}$\;
			}
		}

		\While{$stack$}{
			$\tau \leftarrow stack.\randompop()$ \tcp*[h]{Pop $\tau$ in breadth-first random}\;
			$\ell_{\tau-last} \leftarrow \lastlayout(\tau)$\;
			\eIf{($\ell_{start} \simeq \ell_{\tau-last}$) $\land$ $([e_0, \ldots, e_{i-1}] \leadsto \tau \leadsto [e_i, \ldots, e_n]) \land |\tau| < \maxlen$} {
					\tcp*[h]{Form a loop trace satisfying \texttt{Connect\_with}.}\;
				$candidate\_traces \leftarrow candidate\_traces :: [\tau]$
			} {
				\ForEach{$\langle \ell_{\tau-last}, e, \ell_{next}' \rangle \in M[\ell_{\tau-last}]$}{
					$stack.\push(\tau :: [e])$ 
				}
			}

			\If{$|candidate\_traces| > \maxnum$}{
				break
			}
		}
		\Return $candidate\_traces$
	}
	\caption{Generate mutant tests from a seed test}
	\label{algo:mutant_generation}

\end{algorithm}

\begin{definition}
\noindent\textbf{Independent event trace}. In our context, 
a valid inserted trace $\tau$ should satisfy: $e_1'.r(\ell_{i})$
is \emph{inactive}, \ie, the receiver view of $\tau$'s first event
$e_1'$ on $\ell_i$ is independent from any active views on $\ell_{i}$.
In this case, we say $\tau$ is an independent event trace.
In other words, $\tau$ is independent from subsequent
events in the seed test. Here, we leverage the \activeviewinformation
maintained by $\Pi_{P}(E)=\langle L, I, H \rangle$
in Section~\ref{sec:seed_test}. Section~\ref{sec:fp_analysis}
validates the generality of this assumption.
\end{definition}

\begin{definition}
\noindent\textbf{\texttt{Connect\_with} ($\leadsto$)}. 
Let $\tau_1=[e_1, e_2, \ldots, e_{|\tau_1|}]$ and 
$\tau_2=[e_1', e_2', \ldots, e_{|\tau_2|}']$ be two arbitrary event traces. 
We say $\tau_1$ can connect with $\tau_2$ (denoted 
by $\tau_1 \leadsto \tau_2$) if $\tau_2$'s first event
$e_1'$ can locate a similar receiver view on $\tau_1$'s
last GUI layout $\ell_{|\tau_1|+1}$, \ie, $e_1'.r(\ell_{|\tau_1|+1}) \neq \bot $. 
\end{definition}

Thus, a valid inserted trace $\tau$ should satisfy
$[e_1, \ldots, e_{i-1}] \leadsto \tau \leadsto [e_i, \ldots, e_n]$, \ie,
$\tau$ can connect the event traces before and after
the insertion position, \ie, $e_1'.r(\ell_{i}) \neq \bot \land e_i.r(\ell_{m+1}') \neq \bot$.
Specifically, we will use the \texttt{Similar\_with} relation defined in Section~\ref{sec:seed_test} to locate similar views.

\vspace{2pt}
\noindent\emph{\underline{Example}}. Let Scenario I in 
Fig.~\ref{fig:activity_diary_example}(a) be a seed test.
Assume that we select $\ell_3$  as a pivot layout to insert
independent traces.  The layout $\ell_3$ can be mapped to an abstract
state $s_1$ in the transitional model (Fig.~\ref{fig:activity_diary_example}(c)). 
From $\ell_3$'s
layout information, we know that \emph{Cinema}, \emph{Clearning} and
\emph{Sleeping} are independent views and \emph{Cinema} is
active. Thus, from $\ell_3$, a valid event trace $\tau$ could start from
\emph{Cleaning} or \emph{Sleeping}.  If $\tau$'s
maximum length is set as 2, we can generate at least two valid independent
event traces from the model (denoted by the loop traces in red
lines), \ie,
$\tau_1= \langle Cleaning, Camera \rangle$ , and $\tau_2= \langle Sleeping, Camera\rangle$. 
Note that $\tau_1$ and $\tau_2$ satisfy the \texttt{Connect\_with} relation.
Take $\tau_1$ as an example, $\tau_1$'s first event, \ie, 
clicking \emph{Cleaning}, can find a similar receiver view \emph{Cleaning} on
the runtime layout $\ell_3$ ($\tau_1$'s starting layout), 
while the original event clicking \emph{Nav.} on $\ell_3$ (following $\tau_1$)
can find a similar receiver view \emph{Nav.} on the abstract layout $s_1$
(\ie, $\tau_1$'s last layout). Thus, $\tau_1$ forms a valid independent loop trace that 
can start from $\ell_3$ and return back to $\ell_3$.

\vspace{2pt}
\noindent\textbf{Generate mutant tests from a seed test}.
Algorithm~\ref{algo:mutant_generation} gives how our
approach generates independent event traces to manufacture
property-preserving mutants. The algorithm iteratively
selects each GUI layout of a seed test to generate mutants
(Lines 3-7). It first maps the pivot layout into an abstract
state $\ell_{abs}$ on the transitional model $M$ (Line 4),
from which it starts searching independent traces (Line 5).
In function \texttt{Search\_traces}, the search of valid independent 
event traces is conducted on the transitional model $M$ in a breadth-first
random manner (Lines 14-23). The search process is detailed in Section~\ref{sec:mutant_generation}.
Specifically, a
valid trace $\tau$ should (1) be independent from subsequent
events (Lines 11-13), and (2) form a loop trace at $\ell_{abs}$
that satisfies the \texttt{Connect\_with} relation (Lines 17-18).

\subsection{GUI-Based Oracle Checking}
\label{sec:oracle}

Now, we turn to the difficult challenge of test oracles and automated
oracle checking.  Our key intuition is that, because the inserted
event trace is independent, it would add, but not remove GUI effects
from the execution of the seed test.  If any GUI effect has been
removed, it indicates a likely property violation. We leverage this
intuition to achieve automated oracle checking.

Specifically, we define \emph{GUI effect} as the GUI changes between
two GUI layouts with similar types, which characterizes the effect of
executing the events in between.

\begin{definition}
\noindent\textbf{GUI Effects}.
A GUI test $E$ can yield a sequence of runtime GUI layouts
$L=[\ell_1, \ldots, \ell_{n+1}]$. Given any two GUI
layouts, $\ell_{i}$ and $\ell_{j}$ ($i<j$), satisfying
\texttt{Similar\_layout\_type($\ell_{i}$, $\ell_{j}$)},
their GUI effect, denoted by $\Delta(\ell_i, \ell_j)$,
is equivalent to the differences of GUI views between $\ell_i$
and $\ell_j$. $\Delta(\ell_i, \ell_j)$ characterizes the
effect (\ie, changes of views) of executing the event trace
$[e_{i}, \ldots, e_{j-1}]$ between $\ell_i$ and $\ell_j$. 
Formally,
$\Delta(\ell_i, \ell_j)$ can be represented as:
\begin{equation}\label{eq:gui_view_differences}
\Delta(\ell_i, \ell_j) \equiv (\ell_i \setminus \ell_j) \uplus
(\ell_j \setminus \ell_i)
\end{equation}
where $\setminus$ denotes set difference and $\uplus$ disjoint union.
We use $(\ell_i \setminus \ell_j)$ and $(\ell_j \setminus \ell_i)$ to
model view deletions and additions, respectively. Therefore,
$\Delta(\ell_i, \ell_j)$ captures both deleted and added views from
$\ell_i$ to $\ell_j$, thus the GUI effects of executing the event
trace $[e_{i}, \ldots, e_{j-1}]$ between $\ell_i$ and $\ell_j$.
\end{definition}

In practice, to efficiently compute $\Delta(\ell_i, \ell_j)$, we adapt
the classic tree edit distance
algorithms~\cite{zhang_shasha_algorithm,apted_tree_edit_distance,tree_edit_distance_survey}
to compute the minimal edit operations between the two ordered GUI
trees $T_i$ and $T_j$ (corresponding to $\ell_i$ and $\ell_j$). By
analyzing these edit operations, we can quickly identify the changes
of views, \ie, which views are deleted, added or changed.
More specifically, $\Delta(\ell_i, \ell_j)$ contains three types of
tuples: (1) $(w, \bot) \in \Delta(\ell_i, \ell_j)$ for $w \in
\ell_i$ denotes a view deletion, (2) $(\bot, w) \in \Delta(\ell_i,
\ell_j)$ for $w \in \ell_j$ denotes a view addition, and (3) $(w,
w')$ for $w \in \ell_i$, $w' \in \ell_j$ and $w \neq w'$
denotes a view change.

%
%

\vspace*{2pt}
\noindent\emph{\underline{Example}}. In Fig.~\ref{fig:activity_diary_example}(a),
the GUI effect between $\ell_4$ and $\ell_6$, \ie, $\Delta(\ell_4, \ell_6)$,
is the deletion of a picture under \emph{Cinema}, which characterizes the effect of executing the event trace [``\emph{Pic.}'', ``\emph{Yes}''] in between. 
When we use tree edit distance to do computation, $\Delta(\ell_4, \ell_6)$ can be represented as $\{(pic_{Cinema}, \bot)\}$. Note that $\ell_4$ and $\ell_6$ is comparable because they are of same layout type (\ie, the \emph{Diary} page), while $\ell_3$ and $\ell_4$ is not comparable because they are of different layout types ($\ell_3$ denotes the \emph{main} page of \emph{ActivityDiary}).
Similarly, in Fig.~\ref{fig:activity_diary_example}(b),
the GUI effect between $\ell_4'$ and $\ell_6'$ (in Fig.~\ref{fig:activity_diary_example}(b)) is $\Delta(\ell_4', \ell_6')=\{(pic_{Cinema}, \bot)\}$, while the GUI effect between $\ell_4'$ and $u_6$ is $\Delta'(\ell_4', u_6)=\{(pic_{Cleaning}, \bot)\}$.

Next, we introduce our method for oracle checking based on the definition
of GUI effects.

\begin{definition}
\noindent\textbf{Oracle Violation}.  Given a GUI seed test $E$ with
its GUI layouts $L=[\ell_1, \ell_2, \ldots, \ell_{n+1}]$, and one of
its mutants $E'$ with GUI layouts $L'=[\ell_1', \ell_2', \ldots,
\ell_{n+1}']$, if there exists $(\ell_i, \ell_j) \in L \times L$
such that $\Delta(\ell_i, \ell_j) \nsubset \Delta(\ell_i', \ell_j')$,
\ie, an oracle violation is detected.
\end{definition}

\begin{figure}[h]
        \centering
        \vspace*{-1.0pc}
	\includegraphics[width=0.6\textwidth]{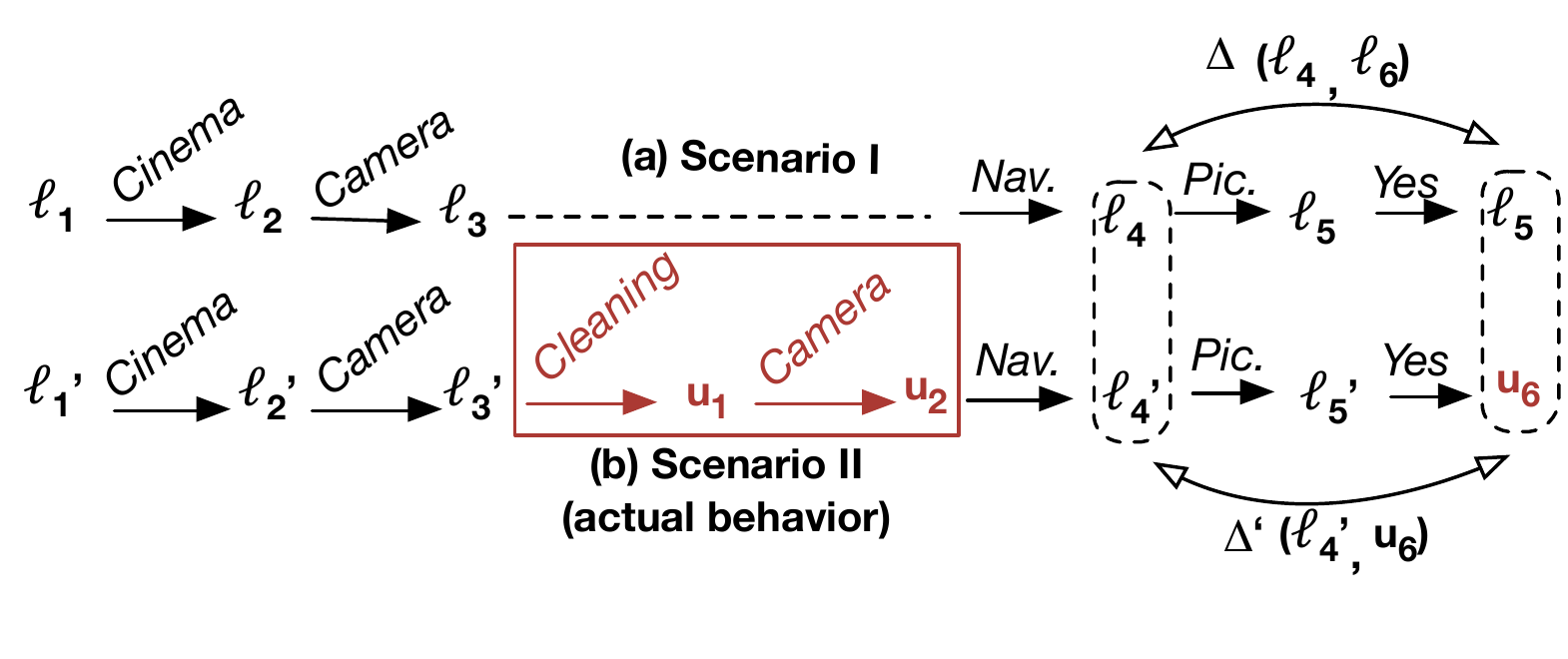}
	\vspace*{-1.7pc}
	\caption{Oracle checking between Scenario I in Fig.~\ref{fig:activity_diary_example}(a) and Scenario II (actual behavior) in Fig.~\ref{fig:activity_diary_example}(b).}
	\label{fig:oracle_checking_diagram}
\end{figure} 


\noindent\emph{\underline{Example}}.
Fig.~\ref{fig:oracle_checking_diagram} shows the oracle checking
between Scenario I in Fig.~\ref{fig:activity_diary_example}(a) and
Scenario II in Fig.~\ref{fig:activity_diary_example}(b) (actual behavior).
The red rectangle annotates the inserted independent events.  
Note that we only need to compare the layouts affected by the inserted traces.
Thus, the only instance of oracle checking in this case is between 
$\Delta(\ell_4, \ell_6)$ from Scenario I
and $\Delta'(\ell_4', u_6)$ from Scenario II because $\ell_4$ and
$\ell_6$ are of same layout type. 
Obviously, due to $\{(pic_{Cinema}, \bot)\} \nsubset \{(pic_{Cleaning}, \bot)\}$, \ie, $\Delta(\ell_4, \ell_6) \nsubset \Delta(\ell_4', u_6)$, a likely bug was reported by
Scenario II (actual behavior).
In another case, if we do oracle checking between Scenario I in Fig.~\ref{fig:activity_diary_example}(a) and Scenario II in Fig.~\ref{fig:activity_diary_example}(b) (expected behavior), 
no oracle violation exists. For example, if we compare 
$\Delta(\ell_4, \ell_6)$ and $\Delta(\ell_4', \ell_6')$, we can easily know 
$\Delta(\ell_4, \ell_6) \subseteq \Delta(\ell_4', \ell_6')$.
Thus, Scenario II in Fig.~\ref{fig:activity_diary_example}(b) (actual behavior) will not report any bug.

\vspace*{2pt}
\noindent\textbf{GUI-based Oracle Checking}.
Algorithm~\ref{algo:oracle_checking} gives our GUI-based oracle
checking procedure. 
$L$ and $L'$ denote the sequences of layouts of a seed test $E$ and one of its mutant test $E'$, respectively. $L_\tau$ is the sequence of layouts due to the inserted trace $\tau$.
Specially, we only need to compare the layouts affected
by the inserted traces to improve checking performance (Line 5), and
focus on the layouts with similar types to improve checking
precision (Line 6).

\begin{algorithm}[t]
	\small
	\SetKwFunction{generate}{Oracle\_Checking}
	\SetKwFunction{similarType}{Similar\_layout\_type}
	\SetKwFunction{true}{True}
	\SetKwFunction{false}{False}
	\SetKwProg{func}{Function}{:}{}
	\func{\generate{Seed test $E$, Mutant test $E'$}}{
		
		$L \leftarrow [\ell_1, \ldots, \ell_k, \ell_{k+1}, \ldots\, \ell_{n+1}]$; $L' \leftarrow [\ell_1, \ldots, \ell_k] :: L_\tau :: [\ell_{k+1}', \ldots\, \ell_{n+1}']$\; 
		
		$O \leftarrow \true$\;
		\ForEach{$(\ell_i, \ell_j) \in L \times L$}{
			\If{$(i \leq k \land k+1 \leq j) \lor (k+1 \leq i < j ) $ }{
				\If{$\similarType(\ell_i, \ell_j)$}{
					\If{$\Delta(\ell_i, \ell_j) \nsubset \Delta(\ell_i', \ell_j')$}{
						$O \leftarrow \false$\;
					}
				}
			}
		}
		\Return $O$\;
	}
	
	\caption{GUI-based Oracle Checking}
	\label{algo:oracle_checking}
\end{algorithm}

\section{Implementation}
\label{sec:impl}


We implemented \tool as an automated, end-to-end functional
fuzzing tool for Android apps. Given an app as input,
it outputs any likely distinct bug reports.
\tool was built upon DroidBot~\cite{LiYGC17} 
and significantly extended Stoat's model mining algorithm~\cite{su_stoat_2017}.
It was written in Python and JavaScript (5,671 lines of
Python for the core algorithms; 1,035 lines of Python 
for parallel fuzzing, and 386 lines of JavaScript
for bug report visualization). Specifically, it uses \texttt{AccessibilityService}~\cite{accessibility_service} to collect
GUI layouts, the \texttt{APTED} algorithm~\cite{apted} to compute tree edit distance, \texttt{ADB}~\cite{adb}
to send GUI events, including \texttt{touch}, \texttt{long-touch}, \texttt{edit}, and \texttt{navigation} (\eg, \texttt{back}, \texttt{scroll}).
We have made \tool publicly available at
\textit{\url{https://github.com/functional-fuzzing-android-apps/home}}.

\subsection{Seed Test Generation}
\label{sec:seed_generation}
We generate random seed tests as the basis of property validation, which
are expected to be much more diverse, practical and scalable to obtain.
To improve the diversity of seeds, we adapted the state coverage-optimized mining
algorithm in Section~\ref{sec:model} for seed generation.
Specifically, we keep the weight information of events during seed test
generation, which will automatically drive each seed test to reach
different GUI pages.
Meanwhile, we exploit the motif events~\cite{mao_sapienz_2016,combodroid}
to improve the chance of generating meaningful seed tests (\eg, more
likely clicking ``OK'' than ``Cancel'' after filling the text fields
in a dialog; more likely taking a picture than clicking ``Cancel''
after opening the camera). 

\subsection{Mutant Generation and Execution}
\label{sec:mutant_generation}
Algorithm~\ref{algo:mutant_generation} uses \texttt{Search\_traces} 
to search independent event traces and create mutants.
However, we face the path explosion problem.
To this end, we did some optimizations.
(1) Algorithm~\ref{algo:mutant_generation} searches the independent traces
in a \emph{breadth-first random search} order. We observe it outperforms depth-first
or pure random search in improving the diversity of mutants according to our preliminary study. 
(2) To avoid path explosion, one independent trace selects the same self-loop
event \emph{at most twice} and \emph{at most three} independent views from
the same group view. 
(3) The transitional model removes redundant events
between any two abstract states to reduce the search space.
Considering the generated mutants are not guaranteedly replayable,
we detect unreplayable mutants at runtime, and use their trace
prefixes to skip similar unreplayable mutants to save testing time.

\subsection{Duplicated Errors and False Positives Reduction}
\label{sec:false_positives}

In practice, \tool may report many of suspicious non-crashing functional
errors, most of which are actually duplicated or false positives.
To reduce \emph{duplicated errors}, we implemented a two-step reduction algorithm.
First, \tool converts each suspicious functional error into a string,
which encodes the witness of oracle violation, \ie, the differences of
GUI views between the seed and its mutant. By string equality comparison,
we remove any redundant functional errors, and only keep distinct ones.
This step \emph{will not} incur any false negatives. Second, inspired by
Engler \etal's ``\emph{bugs as deviant behavior}'' idea ~\cite{engler_sosp01},
\tool ranks these distinct errors according to their occurrences.
The basic idea is that suspicious errors with fewer occurrences are more
likely to be true errors. Specifically, when \tool identifies distinct errors,
it counts the occurrences of them, and then ranks these errors
by their occurrences. In practice, \tool only keeps
the mutants with \emph{1-occurrence} distinct errors (of course, this may unavoidably incur false negatives). 

On the other hand, we observed that many trivial false positives were caused by the app
features and the state abstraction of
GUI model. To this end, we adopted two strategies to remove these trivial false positives.
(1) We replay a seed test twice to identify those GUI views which will automatically change 
by themselves (\eg, time-sensitive text labels, dynamic or non-deterministic app behaviors), 
and exclude these GUI views from oracle checking.
(2) We use a heuristic to decide whether an independent event trace indeed returns back 
to the pivot layout --- if the pivot layout and the layout of ending state of the independent trace
have more than 50\% differences in the texts, we assume these two layouts are likely different.
In other words, this independent event trace fails to return back to the pivot layout and
the corresponding mutant will be excluded.
Section~\ref{sec:bug_finding} shows the reduction effectiveness and
Section~\ref{sec:fp_analysis} discusses the remaining false positives.

We also implemented a bug report visualization tool to ease manual inspection by highlighting critical GUI inconsistencies.
It automatically aligns the seed test and
its mutants side-by-side, and annotates any critical GUI inconsistencies
between different groups of GUI pages. 

\begin{table}[t]
	
	\footnotesize
	\newcommand{\tabincell}[2]{\begin{tabular}{@{}#1@{}}#2\end{tabular}}
	\caption{Challenges faced by \tool, the strategies/heuristics adopted by \tool to tackle the corresponding challenges, and related algorithm pseudocode and functions.}
	\vspace{-0.5pc}
	\begin{tabular}{ |c||c|c|c| }
		\hline
		Challenges &Strategies/Heuristics &Pseudocode &Related Functions \\
		\hline \hline
		Inferring independent views & \tabincell{c}{group view-based inference\\(Section~\ref{sec:seed_test})} & Algorithm~\ref{algo:seed_test_execution} & \tabincell{c}{\texttt{Independent\_from},\\\texttt{Similar\_with},\\\texttt{Similar\_layout\_type}}  \\ \hline
		\tabincell{c}{Generate independent event traces\\(trace space explosion)} & \tabincell{c}{model-based independent\\event trace query\\(Section~\ref{sec:mutant_test},~\ref{sec:mutant_generation})} & Algorithm~\ref{algo:mutant_generation} & \tabincell{c}{\texttt{Equivalent\_with},\\\texttt{Connect\_with},\\\texttt{Search\_traces}}\\
		\hline
		Precise identification of property violation & \tabincell{c}{GUI tree-based \\oracle computation\\(Section~\ref{sec:oracle},~\ref{sec:false_positives})} & Algorithm~\ref{algo:oracle_checking} &\texttt{Similar\_layout\_type}  \\
		\hline
		High-quality, meaningful seed tests & \tabincell{c}{random seed generation \\with motif-events (Section~\ref{sec:seed_generation})} & - & -  \\
		\hline
	\end{tabular}
	\label{tab:challenges_heuristics}
\end{table}

\subsection{Challenges, Trade-offs and Limitations}
\label{sec:trade-off}

Due to the diversity and complexity of apps in practice,
we have to balance between generalizability and precision of our approach.  
For example, specific app features or GUI designs may affect the
precision of inferring independent views and oracle checking.
Thus, \tool adopts some trade-offs when implementing \texttt{Equivalent\_with}, \texttt{Independent\_from}, \texttt{Similar\_with}, \texttt{Connect\_with}, \texttt{Search\_traces} and \texttt{Similar\_layout\_type}.
These trade-offs may lead to imprecise results, and thus incur
false positives or negatives. 
The false positive analysis in Section~\ref{sec:fp_analysis} also reflects
the impacts of these trade-offs. Additionally, \tool's effectiveness is also
limited by the seed test quality and diversity of inserted traces.
Section~\ref{sec:discussion} discusses the quality of seed and mutant tests 
generated by \tool in finding non-crashing bugs.
Table~\ref{tab:challenges_heuristics} summarizes the challenges faced by \tool, the strategies or heuristics adopted by \tool to tackle the corresponding challenges, and the related algorithm pseudocode and functions.

\section{Evaluation}
Our evaluation aims to answer four research questions:

\begin{itemize}[leftmargin=*]
	
	\item \textbf{RQ1 (Bug Finding)}: 
	Can \tool automatically find non-crashing functional bugs in real apps? Can existing fully automated GUI testing tools find such non-crashing bugs?
	
	\item \textbf{{RQ2 (Code Coverage)}}: 
	Can \tool improve code coverage via functional fuzzing?

	\item \textbf{{RQ3 (Oracle Precision)}}: 
	How is \tool's precision in finding non-crashing functional bugs? 
	
	\item \textbf{{RQ4 (Bug Types)}}: 
	What types and characteristics of non-crashing bugs can \tool find?
	
\end{itemize}

\subsection{Evaluation Setup}
\label{sec:evaluation_setup}
\noindent\emph{\textbf{Evaluation Environment}}. 
\tool runs on a 64-bit Ubuntu 18.04 machine (64 cores, AMD 2990WX
CPU, and 128GB RAM) to do functional fuzzing. It runs tests on
Android emulators. Each emulator is configured as a Google Nexus 7
device with 2GB RAM, X86 ABI image (KVM powered), and the Marshmallow
version (SDK 6.0, API level 23). Different types of external
files (including 2 PNGs/2 MP3s/2 PDFs/2 TXTs/1 DOCX) are stored
in the SDCard to facilitate file access from apps.

\noindent\emph{\textbf{App Subjects}}. 
We chose open-source apps in the wild as the subjects 
to evaluate \tool because we can directly communicate with app developers
for bug confirmation.
Specifically, we scrawled all the apps from F-Droid~\cite{fdroid},
the largest open-source Android app market, and also collected the subjects
from recent research on GUI testing of Android ~\cite{choudhary_automated_2015,issta18_yakusu,Amobile18_configuration,fse19_preference_testing}. Among these apps, we selected the apps on GitHub,
by which we can submit bug reports. We obtained 1,752 candidates, and
further filtered apps following the following criteria: 
(1) \emph{Real-world apps} --- The app is officially released on
Google Play~\cite{google_play} for users to download 
and install --- 807 candidates remained; 
(2) \emph{Popular apps} --- The app has 10,000+ installations
on Google Play \emph{and} 100+ stars on GitHub --- 348 apps remained; 
(3) \emph{Well-maintained} --- The app is active and has been maintained for a long time (\eg, two years) since its first release;
we assume such apps are usually stable and well-tested by their
developers (and users) --- 115 candidates remained; and 
(4) \emph{Self-contained apps} ---
Since Android apps are diverse,
for an objective evaluation of \tool, we focus on ``self-contained'' apps, whose
functionalities can be automatically explored without
specific human inputs or participation,
because \tool uses random seed tests.
Thus, we filtered apps that require valid user account
(\eg, email/social media clients), interact with 
other devices (\eg, messaging and IoT apps), require specific
human inputs or gestures (\eg, navigation and gaming apps). 
At the end, we obtained 26 apps that meet these criteria.
\emph{Note that, in principle, \tool is not limited to these apps if provided with proper information (\eg, valid user account or text inputs), which is an orthogonal issue for any automated testing tools.}

\begin{table*}[t]
	\footnotesize
	\newcommand{\tabincell}[2]{\begin{tabular}{@{}#1@{}}#2\end{tabular}}
	\renewcommand{\arraystretch}{1}
	
	\caption{App subjects used in our evaluation. \#L. \#Br., \#M., \#Cl. represent the numbers of code lines, branches, methods and classes, respectively (K=1,000, M=1,000,000).}
	\vspace*{-0.5pc}
	\label{table:app_subjects}
	\centering
	\begin{adjustbox}{max width=\textwidth}
		\begin{tabular}{|c||c|c|c|c|c|c|c|c|c|c|c|c|}
			\hline
			App Name &Feature Description &\tabincell{c}{Version} &\tabincell{c}{First\\Release} &\tabincell{c}{\#Google Play\\Installations} &\tabincell{c}{\#GitHub\\Stars} &\tabincell{c}{\#L.} &\tabincell{c}{\#Br.} & \tabincell{c}{\#M.}  & \tabincell{c}{\#Cl.} \\
			\hline \hline
			\emph{ActivityDiary} &Personal Diary & 1.4.0 &Oct. 2017 & 1K$\sim$5K &56 &4,238 &1,169 & 611 & 157 \\
			\emph{Transistor} &\tabincell{c}{Radio Program\\ Player}  & 3.2.2 &Oct. 2015 & 10K$\sim$50K & 254  &3,024 & 1,003 & 551 & 137 \\
			\emph{Tasks} &\tabincell{c}{Task Manager\\ \& Reminder} & 6.6.5 &Dec. 2008 &100K$\sim$500K &1.3K &31,757 & 8,386 & 6,187 & 984 \\
			\emph{UnitConverter} &Unit Converter & 5.5.1 &Jun. 2014 &1M$\sim$5M &144 &1,525 & 304 & 300 & 88 \\
			\emph{RadioDroid} &\tabincell{c}{Radio \\Streaming Player} & 0.79 &Jan. 2013 &50K$\sim$100K &323 &8,085 & 2,397 & 1,426 & 309 \\
			\emph{SkyTube} &\tabincell{c}{YouTube \\Video Player} & 2.97 &Nov. 2015 &100K$\sim$500K &843 &8,469 & 2,531 & 1,889 & 318 \\
			\emph{SimpleTask} &\tabincell{c}{Cloudless \\To-do Lists} & 10.3.0 &Mar. 2013 &10K$\sim$50K & 390 & 2,512 & 1,022 & 403 & 83 \\
			\emph{Fosdem} &\tabincell{c}{Conference \\Scheduler Browser} & 1.6.2 &Dec. 2013 &10K$\sim$50K & 111 & 5,563 & 1,419 & 1,106 & 290 \\
			\emph{Markor} &\tabincell{c}{Markdown \\Note Editor} & 2.0.1 &Dec. 2014 &50K$\sim$100K & 1.2K & 8,149 & 3,801 & 1,516 & 232 \\
			\emph{AnkiDroid} &\tabincell{c}{Advanced \\Flashcard Learning} & 2.9.6 &Jun. 2009 &5M$\sim$10M & 3K & 29,548 & 11,047 & 4,157 & 543 \\
			\emph{AnyMemo} &\tabincell{c}{Language \& \\Vocabulary Learning} & 10.11.4 &Jan. 2010 &100K$\sim$500K &122 &13,003 & 3,060 & 2,809 & 649 \\
			\emph{MyExpense} &\tabincell{c}{Personal \\Financial Assistant} & 3.0.9.1 &Mar. 2011 &500K$\sim$1M &278 &34,549 & 13,466 & 5,935 & 836 
			\\
			\hline 
		\end{tabular}
	\end{adjustbox}
\end{table*}

To further ensure the diversity of apps, we classified these 26 apps plus the app \emph{ActivityDiary} into four different categories according to 
their main features (\eg, personal management, multimedia), and then
randomly selected 3 apps from each category. Thus, we focus on
the 12 selected apps (including \emph{ActivityDiary}). Table~\ref{table:app_subjects} gives the details of
these subjects. \emph{Feature Description} briefs their functionalities.
\emph{Version} gives the versions of their \emph{latest releases} 
on Google Play at the time of our study, which our evaluation targets.
\emph{First Release} gives the time of their first releases. 
\emph{\#L.}, \emph{\#Br.}, \emph{\#M.} and \emph{\#Cl.} give their
numbers of code lines, branches, methods and classes, respectively. 

\noindent\emph{\textbf{Evaluation Setup}}. 
\tool was configured with 1 hour for mining the GUI transitional model
and 20 random seed tests (each with up to 15 events) for fuzzing.
At each insertion position, \tool generates up to 300 mutants
(one seed test has up to 15 insertion positions);
and each inserted event trace contains up to 8 events.
To scale the testing, \tool runs mutants on 16 emulators
(the maximum number of emulators allowed by Android system) in parallel
on one machine. Under these settings, it took around 25$\sim$35
machine hours to finish testing one app 
(including the four core steps and bug report reduction).
The whole evaluation took roughly 18 machine days in total for the twelve app subjects.
Note that the fuzzing cost can be significantly reduced by using more machines, Android devices, and parallel threads.
In addition, we selected two state-of-the-art/practice fully automated GUI testing tools for Android, \textsc{Monkey}~\cite{monkey} and \textsc{Ape}~\cite{ape_icse2019}, to compare them with \tool and verify whether they can find non-crashing bugs. We allocated 30 hours for the two tools to test each of the twelve app subjects, and inspected the found bugs.

\subsection{RQ1: Bug Finding}
\label{sec:bug_finding}

Table~\ref{tab:bugs_overview} shows the statistics of bugs found by \tool. 
\tool found \totalbugs non-crashing functional
bugs, all of which were confirmed and 22 were
already fixed. As a by-product of our testing,
26 crash bugs were also discovered with 24 confirmed
and 14 fixed. In total, \tool found 60 bugs with
58 confirmed and 36 fixed. All these bugs
were discovered in the latest app releases at the time of our study
and previously unknown. 
Section~\ref{sec:bug_types} discusses the types of the 34 non-crashing bugs.

Table~\ref{tab:bugs_comparison} shows the comparison results.
We can see that \textsc{Monkey} and \textsc{Ape} \emph{cannot} detect any of the 34 functional bugs because they are limited to finding crash bugs. Specifically,
we find \tool and \textsc{Monkey} have no common crash bugs, while \tool and \textsc{Ape} have 6 common crash bugs.
These results clearly show \tool complements existing fully automated testing for crash bugs.

\begin{table}[t]
	\begin{minipage}{0.45\linewidth}	
		\footnotesize
		\newcommand{\tabincell}[2]{\begin{tabular}{@{}#1@{}}#2\end{tabular}}
		\caption{Bugs that were found by \tool in the evaluated app subjects.}
		\vspace{-0.5pc}
		\begin{tabular}{ |c||c|c|c| }
			\hline
			Bug Type &\#Reported &\#Confirmed &\#Fixed  \\
			\hline \hline
			\#Non-crashing Bugs &\totalbugs &34 &22  \\ 
			\#Crash Bugs &26 &24 &14  \\
			\hline
			\#Total &60 &58 &36 \\
			\hline
		\end{tabular}
		\label{tab:bugs_overview}
	\end{minipage}\hfill
	\begin{minipage}{0.45\linewidth}
		\footnotesize
		\newcommand{\tabincell}[2]{\begin{tabular}{@{}#1@{}}#2\end{tabular}}
		\caption{Comparison between \tool, \textsc{Monkey} and \textsc{APE} on detecting non-crashing bugs.}
		\vspace{-0.5pc}
		\begin{tabular}{|c||c|c|c|}
			\hline
			Bug Type &\tool  & \textsc{Monkey} &\textsc{Ape}  \\
			\hline \hline
			\#Non-crashing Bugs & \textbf{34} & \textbf{0} & \textbf{0} \\
			\#Crash Bugs & 26  & 25 & 36 \\ \hline
			\#Total & 60 & 25 & 36 \\ \hline
		\end{tabular}
		\label{tab:bugs_comparison}
	\end{minipage}
\end{table}

\begin{table*}[t]
	\footnotesize
	\newcommand{\tabincell}[2]{\begin{tabular}{@{}#1@{}}#2\end{tabular}}
	\renewcommand{\arraystretch}{1}
	
	\caption{Detailed statistics of functional fuzzing on the app subjects. D. E. (Distinct Error), 1-O. D. E. (1-Occurrence Distinct Error), T. P. (True Positive).}
	\vspace*{-0.5pc}
	\label{table:detailed_testing_statistics}
	\centering
	\begin{adjustbox}{max width=\textwidth}
		\begin{tabular}{|c||c|c|c|c|c|c|c|c|c|c|c|c|}
			\hline
			App Name &\tabincell{c}{\#Generated\\Mutants} &\tabincell{c}{\#Executed\\Mutants} & \tabincell{c}{\#Error\\ Mutants} & \tabincell{c}{\#D. E.\\Mutants} & \tabincell{c}{\#1-O. D. E.\\Mutants} & \tabincell{c}{\#1-O. D. E.\\Mutants (Refined)} & \tabincell{c}{\#T. P.\\ Mutants} &\tabincell{c}{\#Distinct \\Non-crashing Bugs} & \tabincell{c}{\#Distinct \\Crash Bugs}  \\
			\hline \hline
			\emph{ActivityDiary} &45,258 & 27,433 &13,822 &1,264 &249 &69 &31 & 7 & 3 \\
			\emph{Transistor} &84,029 & 6,740 &2,497 &194 &46 &16 &12 & 5 & 0 \\
			\emph{Tasks} &37,339 & 33,786 &7,367 &396 &87 &17 &12 & 2 & 2 \\
			\emph{UnitConverter} &73,203 & 21,511 &14,864 &4,015 &229 &106 &50 & 2 & 1 \\
			\emph{RadioDroid} &50,421 & 28,954 &10,538 &819  &95 &62 &19 & 6 & 5 \\
			\emph{SkyTube} &73,853 & 12,525 &7,810 &2,584  &558 &58 &27 & 1 & 8 \\
			\emph{SimpleTask} &59,081 & 13,526 &3,277 &363  &75 &16 &13 & 1 & 0 \\ 
			\emph{Fosdem} &36,594 & 31,990 &6,243 &845  &192 &71 &30 & 1 & 1 \\ 
			\emph{Markor} &57,065 & 20,516 &7,978 &665  &132 &73  &33 & 4 & 1 \\ 
			\emph{AnkiDroid} &59,740 & 38,475 &3,759 &449  &58 &23 &5 & 1 & 2 \\ 
			\emph{AnyMemo} &39,288 & 19,552 &5,255 &448  &95 &83 &22 & 3 & 3 \\
			\emph{MyExpense} &68,565 & 40,925 &9,262 &2,815  &378 &41 &6 & 1 & 0 \\ 
			\hline
			\textbf{Total} &684,436  &295,934  &92,672 &14,857 &2,194 &635 &260  &34 &26 \\
			\hline 
		\end{tabular}
	\end{adjustbox}
\end{table*}

Table~\ref{table:detailed_testing_statistics} gives the detailed
statistics. \emph{\#Generated Mutants}
and \emph{\#Executed Mutants} denote the total number of generated and
executed mutants, respectively. 
\emph{\#Error Mutants}, \emph{\#D. E. Mutants}, and \emph{\#1-O. D. E. Mutants}
denote the total number of error mutants (\ie, suspicious functional errors), 
the distinct error mutants (including 1-occurrence, 2-occurrence, 3-occurrence, \etc), 
the 1-occurrence distinct error mutants (only appear once), respectively.
\emph{\#1-O. D. E. Mutants (refined)} denotes the number of 1-occurrence distinct error mutants after automatically removing trivial false positives (discussed in Section~\ref{sec:false_positives}), which are the error mutants for manual inspection.
\emph{\#T. P. Mutants} denotes the total number
of true positives (\ie, real functional bugs). \emph{\#Distinct Non-crashing Bugs}
and \emph{\#Distinct Crash Bugs} denote the number of distinct non-crashing
bugs and crash bugs, respectively.

Table~\ref{table:detailed_testing_statistics} shows \tool is able
to discover non-crashing functional bugs in all the subjects. Specifically,
for several
highly popular apps, \ie, \emph{Tasks}, \emph{UnitConverter}, \emph{Skytube},
\emph{Markor} and \emph{AnkiDroid},
although they have many developer-written tests and oracles (released with their source code) and large user installations (1M$\sim$10M)/high number of stars (843$\sim$3K),
\tool is still able to find 24 bugs (10 non-crashing bugs and 14 crash bugs) in them. It shows \tool's effectiveness.

From Table~\ref{table:detailed_testing_statistics}, we can see 
{295,934} out of {684,436} ({43.2\%}) 
mutants were executable. The reason is that GUI tests are event
order sensitive. Any mutation-based testing technique like
ours may generate unreplayable tests. Because the inserted event
traces may lead to the GUI states that cannot connect with 
subsequent events. For example, \emph{Transistor} has many
unreplayable mutants due to this reason. 
Additionally, we can see \tool's duplicated errors
reduction is effective, which reduces {92,672} error mutants
to {14,857} distinct ones ({6.2X} reduction rate);
by ranking these distinct error mutants and automatically reducing 
trivial false positives, \tool further constrains
our focus on 635 mutants for manual inspection. 
Section~\ref{sec:fp_analysis} will discuss the false positives and show that the manual inspection cost is acceptable.

\subsection{RQ2: Code Coverage}
\label{sec:code_coverage}
Table~\ref{table:code_coverage} shows the code coverage of all subjects achieved by the 20 random seed tests and their corresponding mutant tests, respectively. The seed tests and the mutant tests are all generated by \tool. 
The main purpose of this measurement is to investigate whether \tool can improve code coverage via functional fuzzing.
Here, the code coverage data is measured by Jacoco~\cite{jacoco} at runtime.
For example, for \emph{ActivityDiary}, the mutant tests generated by \tool achieves 19.1\% and 14.4\% higher line and branch coverage than the seed tests, respectively.
We can see that, \tool on average improved line coverage by {8.9\%}, branch coverage by {10.0\%}, and method coverage by {6.5\%} by generating the mutant
tests from the seed tests.
This coverage improvement is significant, considering achieving high code coverage of Android apps is challenging~\cite{ZhengLLZZDLYX17}.
These results also explain \tool's effectiveness in finding functional bugs.
For example, in 8 apps, the improvements of branch coverage are even higher than those of line coverage.
The improvement of branch coverage indicates the control-flows or even data-flows were more stress-tested by the mutants than the seeds.

\begin{table*}[t]
	\footnotesize
	\newcommand{\tabincell}[2]{\begin{tabular}{@{}#1@{}}#2\end{tabular}}
	\renewcommand{\arraystretch}{1}
	
	\caption{Code coverage of seed tests and their corresponding mutant tests generated by \tool for the 12 app subjects. Line, Branch, Method represent line, branch and method coverage, respectively. The numbers in the brackets represent the coverage improvement of the mutant tests \wrt the seed tests in terms of line, branch and method coverage, respectively.}
	\vspace*{-0.5pc}
	\label{table:code_coverage}
	\centering
	\begin{tabular}{|c||c|c|c|c|c|c|c|c|c|}
		\hline
		App Name &\tabincell{c}{Line (\%)\\(seeds)} &\tabincell{c}{Line (\%)\\(mutants)} & \tabincell{c}{Branch (\%)\\(seeds)} & \tabincell{c}{Branch (\%)\\ (mutants)} & \tabincell{c}{Method (\%)\\(seeds)}& \tabincell{c}{Method (\%)\\(mutants)}  \\
		\hline \hline
		\emph{ActivityDiary} &45.9 & 65.0 (+19.1) &33.8 &48.2 (+14.4) &51.7 &70.0 (+18.3)  \\
		\emph{Transistor} &58.5 & 63.1 (+4.6) &39.1 &43.9 (+4.8) & 61.0 & 65.5 (+4.5)  \\
		\emph{Tasks} &28.6 & 33.9 (+5.3) &16.8 &23.3 (+6.5) &36.4 &40.9 (+4.5)\\
		\emph{UnitConverter} &56.9 &71.6 (+14.7) &32.5 &61.5 (+29) &63.3 &73.7 (+10.4)  \\
		\emph{RadioDroid} &46.6 &53.2 (+6.6) &31.9 &40.0 (+8.1) &48.8 &55.2 (+6.4) \\
		\emph{SkyTube} &44.7 &51.9 (+7.2) &30.5 &37.3 (+6.8)  &50.0 &55.9 (+5.9) \\
		\emph{SimpleTask} &31.3 & 33.7 (+2.4) &15.3 &19.3 (+4.0) &31.0 &34.7 (+3.7) \\ 
		\emph{Fosdem} &60.0 &62.0 (+2.0) &39.7 &44.4 (+4.7) &69.0 &70.3 (+1.3) \\ 
		\emph{Markor} &49.7 &57.7 (+8.0) &34.6 &43.8 (+9.2) &54.4 &60.0 (+5.6)  \\ 
		\emph{AnkiDroid} &32.7 & 42.1 (+9.4) &20.8 &28.0 (+7.2) &41.0 &51.4 (+10.4)  \\ 
		\emph{AnyMemo} &49.3 &53.5 (+4.2) &32.1 &38.7 (+6.6) &55.6 &58.3 (+2.7) \\
		\emph{MyExpense} &25.1 &29.3 (+3.5) &15.6 &19.2 (+3.6) &32.3 &36.6 (+4.3)  \\
		\hline
		\textbf{Average} &\textbf{44.1}  &\textbf{53.0 (+8.9)} &\textbf{28.6} &\textbf{38.6 (+10.0)} &\textbf{49.5} &\textbf{56.0 (+6.5)}  \\
		\hline 
	\end{tabular}
\end{table*}

\subsection{RQ3: Oracle Precision}
\label{sec:fp_analysis}


As shown in Table~\ref{table:detailed_testing_statistics}, 
\tool may report false positives (FPs). 
To this end, we carefully examined all the reported FPs 
to identify the root causes and
understand the oracle precision (and other possible issues).
Table~\ref{table:fp_analysis} gives the detailed FP analysis results. \emph{\#F. P. Mutants} denotes the total number of FP mutants (computed by \emph{\#1-O. D. E. Mutants (refined)} and \emph{\#T. P. Mutants} in Table~\ref{table:detailed_testing_statistics}). 
Specifically, we identified three classes of FPs. 

\emph{One class of FPs is related to the violation of independent view property}.
Some FPs (see \emph{\#I} in Table~\ref{table:fp_analysis}) appear because
the inserted event traces do not satisfy the independent view property. 
The main reason is that
\tool only leverages GUI layouts to automatically 
infer independent views, which sometimes may not be precise.
Fig.~\ref{fig:example_fps} (a) shows an example of \emph{UnitConverter}. 
\tool infers the \texttt{FAB} button (annotated by a red box) is independent from other views, and thus generates a number of independent event traces that start with a \texttt{click} on this button. However, clicking this button will exchange two units between the columns ``From'' and ``To'', which breaks the independent property and affects subsequent events. Fig.~\ref{fig:example_fps} (b) shows another example of \emph{AnyMemo}, where \tool assumes the \emph{shuffle} button (the left red box) and the \emph{repeat} button (the right red box) are independent within their group view. However, they affect each other when both are enabled.



The other two classes of FPs are caused by some open technical challenges.
Some FPs (see \emph{\#L}) appear because \tool sometimes
fails to precisely infer layout types.
\tool takes two comparable GUI layouts (with similar functionality) 
and computes GUI effects. 
However, automatically inferring layout types is difficult.
For example, \tool cannot differ \emph{SkyTube}'s page ``Featured'' (see Fig.~\ref{fig:example_fps} (c)) and
page ``Bookmarks'' (see Fig.~\ref{fig:example_fps} (d)), although we humans can easily tell they are different, which should not be compared. \emph{Fosdem} and \emph{RadioDroid} have similar FPs.

Some FPs (see \emph{\#S}) are caused by \emph{model state abstraction}.
\tool queries independent event traces via the abstract GUI model. This model allows us to generate more interesting mutants because similar states are abstracted to form a lot of loop traces. However, this may incur invalid independent traces in practice, \ie, the layout of ending state of an independent trace and the pivot layout cannot form a \emph{connected, loop trace} at concrete state level. 
For example, in \emph{ActivityDiary}, an invalid trace may start from page Fig.~\ref{fig:example_fps} (e) and ends at page Fig.~\ref{fig:example_fps} (f) because these two pages are similar at the abstraction level (because all the texts are ignored in the abstracted model). The second heuristic we used in Section~\ref{sec:false_positives} to reduce false positives cannot exclude this invalid event trace because these two pages are only different at the page title (``New Activity'' \emph{v.s.} ``Gardening''). In fact, eliminating such cases depends on the precision of inferring layout types at the concrete state level.



\begin{table}[t]
	\begin{minipage}{0.4\linewidth}
	\footnotesize
	\newcommand{\tabincell}[2]{\begin{tabular}{@{}#1@{}}#2\end{tabular}}
	\renewcommand{\arraystretch}{1}
	
	\caption{Detailed analysis results on false positives (FPs).}
	\label{table:fp_analysis}
	\centering
	\begin{adjustbox}{max width=\textwidth}
		\begin{tabular}{|c||c||c|c|c|c|c|c|c|c|}
			\hline
			App Name &\tabincell{c}{\#F. P.\\ Mutants} & \tabincell{c}{\#I} & \tabincell{c}{\#L} & \tabincell{c}{\#S} \\
			\hline \hline
			\emph{ActivityDiary} &38 & 31 &5 &2   \\
			\emph{Transistor} &4 & 4 & 0 & 0  \\
			\emph{Tasks} &5 & 5 &0 &0  \\
			\emph{UnitConverter} &56 &50  &6 &0   \\
			\emph{RadioDroid} &43 &8  &10 &25   \\
			\emph{SkyTube} &31 &16  &15  &0   \\
			\emph{SimpleTask} &3  & 2 &1  &0   \\ 
			\emph{Fosdem}&41  & 0 &22 &19   \\ 
			\emph{Markor}  &40 &7  &5  &28   \\ 
			\emph{AnkiDroid} &18 & 16   &0  &2   \\ 
			\emph{AnyMemo} &61 &34  &12  &15   \\
			\emph{MyExpense} &35 &0 &0  &35   \\ 
			\hline
			\textbf{Total} &375  &173 &76  &126  \\
			\hline 
		\end{tabular}
	\end{adjustbox}
	\end{minipage}
	\begin{minipage}{0.55\linewidth}
		\centering
		\includegraphics[width=70mm]{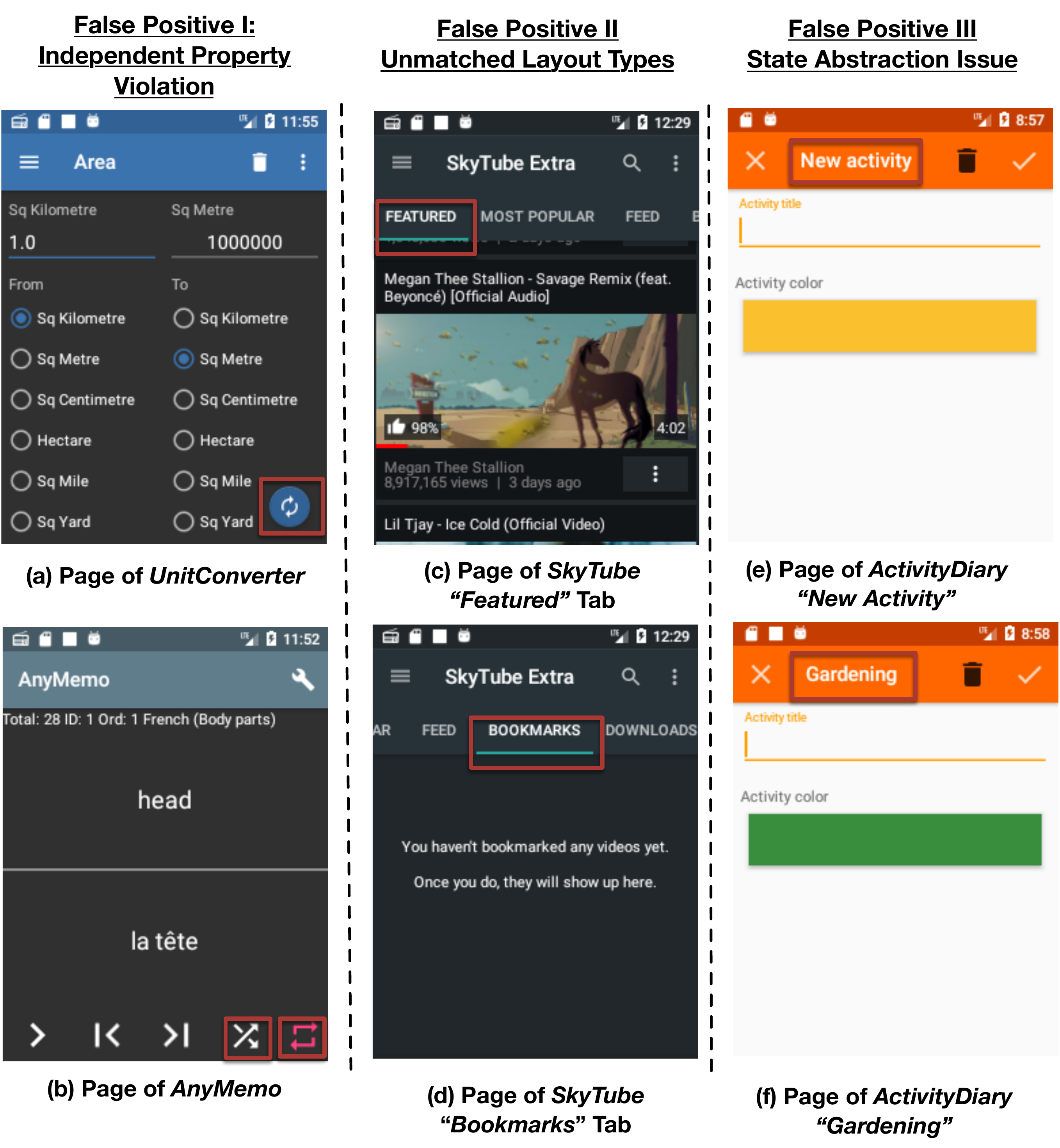}
		\captionof{figure}{Examples of different false positive types. The red boxes annotate the root causes.}
		\label{fig:example_fps}
	\end{minipage}\hfill
\end{table}

\vspace{1pt}
\noindent{\textbf{Discussions}}.
According to Table~\ref{table:fp_analysis}, \tool's true positive rate is 40.9\% (260/635). 
It is worth noting that this is the result in a very challenging problem setting (\ie, finding non-crashing functional bugs without any human-provided tests and oracles).
Existing fully automated tools cannot find the \totalbugs non-crashing bugs. 
To put things into perspective, high-impact, well-tuned industrial bug finding tools like Facebook's \textsc{Infer} was found to have a 26.3\% true positive rate for detecting null pointer dereferences, an extensively studied problem which Infer targets, by a recent study~\cite{WangWGW20}.
Also, the FP analysis shows some FPs could be further automatically removed
with light developer input (\eg, providing information on layout types, independent views and tab types). 
Thus, the current result is already strong for the problem setting.
On the other hand, the benefits of \tool in finding non-crashing bugs
can outweigh the manual cost of inspecting these FPs.
First, manual validation (\eg, writing tests and oracles) is the only alternative 
to discover these bugs, but is more costly, small-scale, 
and often ineffective in finding bugs.
For example, 8 out of the 12 apps in fact have manually-written tests and oracles, which did not reveal these bugs however.
Second, our experience shows that manually checking one suspicious mutant takes roughly 
half a minute. Thus, the manual cost is less than one hour for each app
(see \emph{\#1-O. D. E. Mutants (refined)} in Table~\ref{table:detailed_testing_statistics}).
This is a low, affordable cost since it is much more expensive to manually 
write and maintain GUI tests.
Moreover, in the context of classic static analysis, to check whether an alert is indeed a bug or not, one typically has to look at the code and its context and try to decide, which is tough. While in our context, one only needs to look at the GUI pages on the test executions, which is much easier. 




\subsection{RQ4: Bug Types and Assorted Samples}
\label{sec:bug_types}
The non-crashing functional bugs found by \tool are diverse. 
According to the bug manifestations and consequences, we are able to categorize
the \totalbugs non-crashing bugs into six main types (summarized
in Table~\ref{table:bug_types}).
We illustrate these six bug types with some assorted bug samples.

\begin{figure*}[t]
	\centering
	\includegraphics[width=1.0\textwidth]{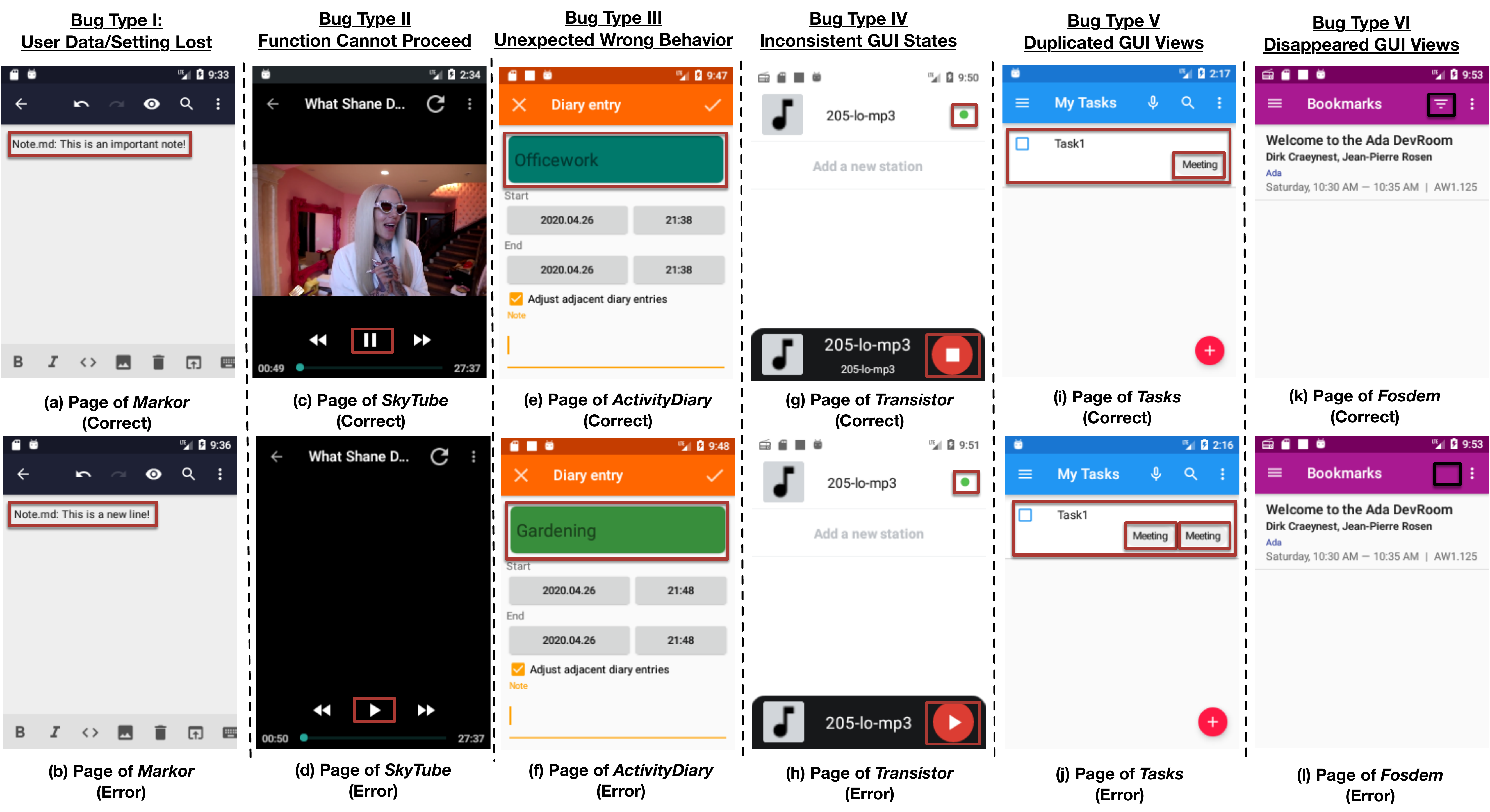}
	\vspace*{-1.7pc}
	\caption{Examples of different functional bug types. In each group, the page at top shows the correct behavior, while the page at bottom shows the erroneous behavior, and the red boxes indicate the clues of each issue. }
	\label{fig:example_bugs}
\end{figure*}

\begin{table}[t]
	\footnotesize
	\newcommand{\tabincell}[2]{\begin{tabular}{@{}#1@{}}#2\end{tabular}}
	\renewcommand{\arraystretch}{1}
	
	\caption{Different types of non-crashing functional bugs found by \tool.}
	\label{table:bug_types}
	\vspace*{-0.5pc}
	\centering
	\begin{adjustbox}{max width=\textwidth}
		\begin{tabular}{|c||c|c|}
			\hline
			Type ID & Bug Type & \tabincell{c}{\#Bugs}  \\
			\hline \hline
			\textbf{T1} & user data/setting lost & 4   \\
			\textbf{T2} & function cannot proceed & 7  \\
			\textbf{T3} & unexpected wrong behavior & 7  \\
			\textbf{T4} & inconsistent GUI states &6    \\
			\textbf{T5} & incorrectly displayed GUI views &7    \\
			\textbf{T6} & GUI design issue &3   \\
			\hline
		\end{tabular}
	\end{adjustbox}
\end{table}

\vspace{1pt}
\noindent{\emph{\textbf{T1: User data/setting lost}}}. This bug type leads to user data or setting lost. Sometimes, this bug type can bring severe consequences and critical user complaints. For example, \emph{Markor}, a markdown note editor, ever had a data loss issue. One user critically complains ``\emph{Markor just went from hero to zero, in my books, as a file I had been working on just disappeared, without a trace! [...] I don't feel like I can currently trust markor with anything important, like the file I was just working on.}''.
\tool uncovered a data loss issue in \emph{Markor}, a markdown note editor.
The original content of a note file will be silently overwritten if the user creates another note file with the same name. Fig.~\ref{fig:example_bugs}(a) and Fig.~\ref{fig:example_bugs}(b), respectively, show the main page where a user can create two notes with the same name and the latter note's content overwrites that of the former one. The developer quickly confirmed and fixed this bug.

\vspace{1pt}
\noindent{\emph{\textbf{T2: Function cannot proceed}}}. This bug type means one specific app function that works well before suddenly cannot proceed anymore and loses effect.
\tool found such an issue in \emph{SkyTube}, a YouTube video player.
Originally, a user can arbitrarily pause and play any video at his/her will.
However, if the user opens the menu, selects an option that plays the video in a browser, and returns back, he/she cannot play the video anymore --- clicking the play button cannot load the video; the play button cannot change to a pause button as well. After we reported this issue, the developer quickly confirmed and fixed it with positive feedback ``\emph{Thanks for the excellent bug report, it was very helpful to fix the bug!}``. 

\noindent{\emph{\textbf{T3: Unexpected wrong behavior}}}.
This bug type means the specific functionality shows wrong behavior \wrt its previous correct behavior, \eg, an incorrect GUI page was erroneously opened, a deleted GUI view was suddenly restored. \tool found such an issue in \emph{ActivityDiary}. Fig.~\ref{fig:example_bugs}(e) shows the history detail page of \emph{Officework}, which was correctly opened after the user (1) starts \emph{Officework} from the main page, and (2) clicks its statistics region. However, if the user does three additional actions between (1) and (2) --- rename \emph{Officework} to another arbitrary name (say \emph{Shopping}); start another activity (say \emph{Gardening}); and undo \emph{Gardening} (\ie, return back to \emph{Shopping}). \emph{ActivityDiary} will erroneously open the history page of \emph{Gardening} rather than that of \emph{Shopping} (show in Fig.~\ref{fig:example_bugs}(f)). 
If at this time the user modifies the history data of \emph{Gardening}, it will break data consistency. 
The developer commented ``\emph{(This is) definitely a bug, but I have to dig deeper in the code to get an idea how could it go wrong}''.

\vspace{1pt}
\noindent{\emph{\textbf{T4: Inconsistent GUI states}}}.
This bug type means the GUI states are inconsistent for specific functionality, which counters users' intuition on app function.
\tool found such an issue in \emph{Transistor}, a radio program player.
In the normal scenario,
Fig.~\ref{fig:example_bugs}(g) shows the page where a radio program is playing, and meanwhile a green playing indicator is shown near the program name. When the user stops the program, this green indicator will disappear. However, if the user renames the program when it is playing, and then stops it. The playing button will correctly change to a pause button, but the green playing indicator keeps there and indicates the program is playing (see Fig.~\ref{fig:example_bugs}(h)). This issue confuses users' intuition on the play-stop function. In fact, at the same time, the new program name will get lost.  
The developer was quite annoyed by this issue and commented ``\emph{This [...] related to the way Transistor handles stations internally. The root (cause) is so deep in the app's architecture, [...], it needs to wait for the big Kotlin re-write. Without a rewrite this would end in a bunch of nasty hacks.}''. Finally, the developer re-wrote the architecture, and fixed this issue.

\vspace{1pt}
\noindent{\emph{\textbf{T5: Incorrectly displayed GUI views}}}.
Some GUI views are erroneously duplicated.
\tool found such an issue in \emph{Tasks}, a task manager \& reminder.
Fig.~\ref{fig:example_bugs}(i) shows its main page when \emph{Task 1} was successfully created with a tag \emph{Meeting}. However, when this task was additionally inserted with two attachments (\eg, two pictures). \emph{Task 1} will be erroneously associated with two duplicated \emph{Meeting} tags (see Fig.~\ref{fig:example_bugs}(j)).

Some GUI views inadvertently disappear.
\tool found such an issue in \emph{Fosdem}, a conference schedule browser.
Fig.~\ref{fig:example_bugs}(k) shows the correct \emph{Bookmarks} page. However, if the user (1) opens the menu (the three dots on the top right corner), (2) does two consecutive selections on the option ``Search events'', and (3) presses back, the filter option on the tool bar inadvertently disappears (Fig.~\ref{fig:example_bugs}(l)). Only restarting the app or reopening the page can recover this option view.

Some views are incorrectly displayed. 
For example, \tool found such an issue in \emph{RadioDroid}. Due to this issue, all page titles are displayed as ``Setting'' if the user changes an irrelevant setting option. This option only alters the location of navigation menus between the page bottom and the navigation drawer.

\vspace{1pt}
\noindent{\emph{\textbf{T6: GUI design issues}}}.
This bug type is related to the issue of functional GUI design, which
could confuse users and disrupt user experience.
For example, \tool found such an issue in \emph{RadioDroid}. 
The issue manifests itself as the \emph{history} page fails to record some ever played radio programs. After the discussion with developers, we know that \emph{RadioDroid} maintains two types of history data, \ie, \emph{history} and \emph{track history}. The \emph{track history} only records those radio programs with track information. However, \emph{RadioDroid} has never clearly differed these two types of history data on its GUIs (use the word \emph{history} for both functional pages). Finally, developers confirmed this issue and reworked the GUI design to avoid user confusion.


\begin{table}
	\begin{minipage}{0.45\linewidth}
		\centering
		\includegraphics[width=61mm]{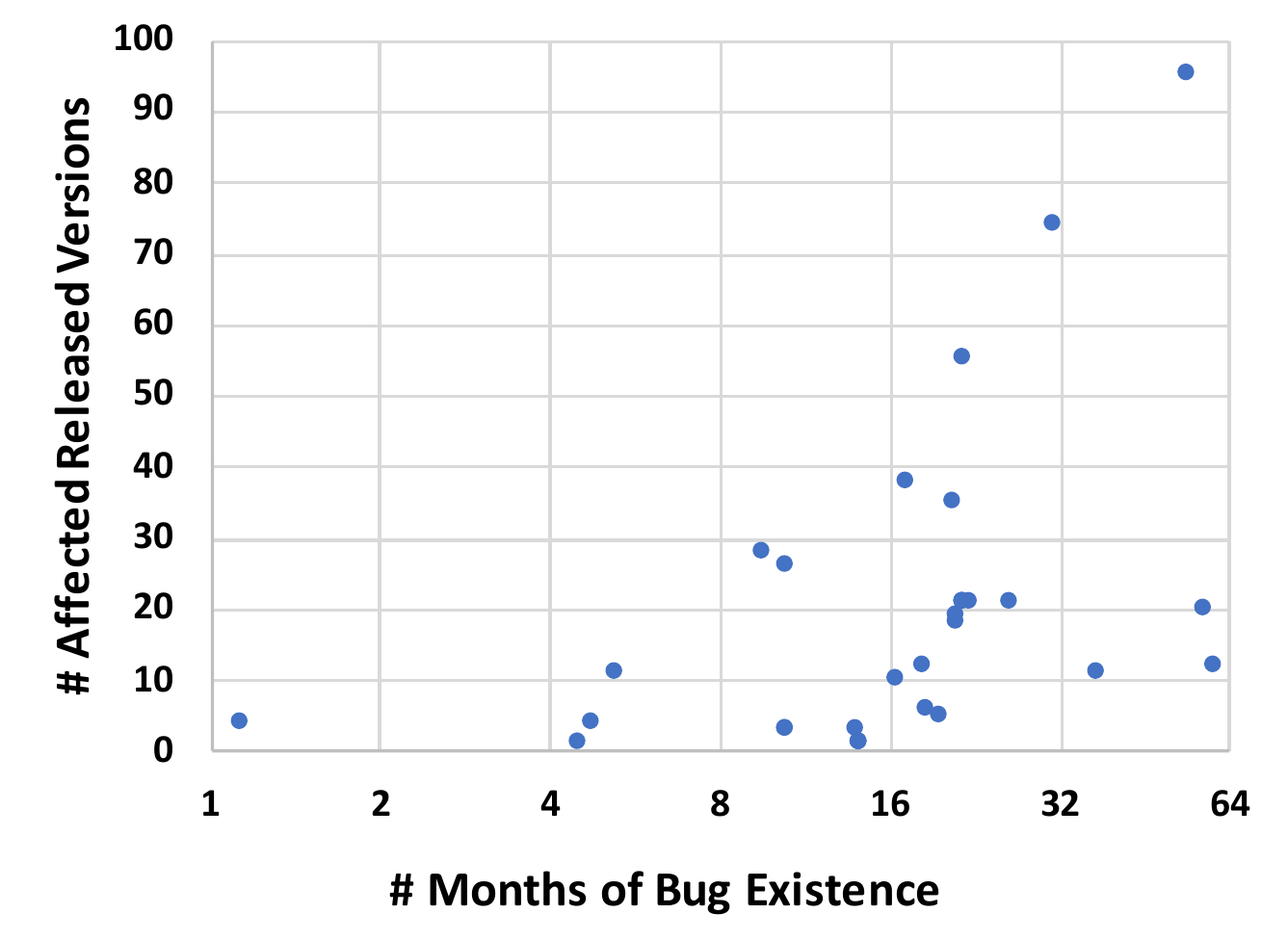}
		\vspace{-2mm}
		\captionof{figure}{Statistics of \totalbugs confirmed functional bugs: (1) \#months they have resided (x-axis, in logarithmic scale); (2) \#affected releases (y-axis), before they were uncovered.}
		\label{fig:bug_quality}
	\end{minipage}\hfill
	\begin{minipage}{0.5\linewidth}
		\footnotesize
		\newcommand{\tabincell}[2]{\begin{tabular}{@{}#1@{}}#2\end{tabular}}
		\renewcommand{\arraystretch}{1}
		
		\caption{Quality of seed and mutant tests generated by \tool. 1-O. D. E. (1-Occurrence Distinct Error), T. P. (True Positive), Insert. Pos. (Insert Position).}
		\vspace{-2mm}
		\label{table:mutation_ability}
		\centering
		\begin{adjustbox}{max width=\textwidth}
			\begin{tabular}{|c||c|c|c|c|c|c|c|}
				\hline
				App Name &\tabincell{c}{\#1-O. D. E. \\Seeds (refined) (A)} &\tabincell{c}{\#T. P.\\Seeds (B)} & \tabincell{c}{B/A} & \tabincell{c}{T. P. Insert.\\ Pos.  (\%)}   \\
				\hline \hline
				\emph{ActivityDiary} &15 & 8 &53.3\% & 30.4\%   \\
				\emph{Transistor} &13 & 5 &38.5\% & 73.3\%   \\
				\emph{Tasks} &10 & 7 &70.0\% &45.7\% \\
				\emph{UnitConverter} &8 &2 &25.0\% & 35.0\%   \\
				\emph{RadioDroid} & 11 & 5 & 45.0\% & 63.9\%  \\
				\emph{SkyTube} & 15 & 2 & 13.3\% & 23.0\%   \\
				\emph{SimpleTask} & 7 & 1 & 14.3\% & 25.0\%  \\ 
				\emph{Fosdem} &9 &1 &11.1\% &60.0\%  \\ 
				\emph{Markor} &11 &4 &36.6\% &52.5\%   \\ 
				\emph{AnkiDroid} &14 & 1 & 7.1\% &20.4\%    \\ 
				\emph{AnyMemo} &12 & 4 & 33.3\% & 48.7\%  \\
				\emph{MyExpense} &15 &2 &13.3\% &18.6\%   \\ 
				\hline
				\textbf{Total} &140  &42 &- &-   \\
				\hline 
			\end{tabular}
		\end{adjustbox}
	\end{minipage}
\end{table}

\subsection{Discussion}
\label{sec:discussion}

\noindent\textbf{The non-crashing functional bugs found by \tool are nontrivial and long latent}. 
Fig.~\ref{fig:bug_quality} characterizes the \totalbugs non-crashing bugs. 
Each point denotes
one bug, which is characterized by (1) \#months it has resided in the app and (2) \#affected releases on Google Play before uncovered by \tool (some points are overlapped). We can see {26} out of the \totalbugs bugs (78.8\%) were unnoticed for more than one year, and {19} out of the \totalbugs bugs (57.6\%) affected more than ten releases.
For example, \tool uncovered a user setting lost issue in \emph{UnitConverter} (1M$\sim$5M installations), which leads to an incorrect group separator shown in the converted values. This issue was unnoticed for more than four years and its manifestation requires a minimal 10-event GUI test.
\tool also found an inconsistent GUI states issue in \emph{Markor} ({1.2K} GitHub Stars), which leads to \texttt{Undo} and \texttt{Redo} behave inconsistently when editing different lengths of texts. This issue was introduced by another bug fix three years ago, and resides in 74 releases. Now both bugs were fixed.
These results show many found bugs are non-trivial and long latent.
Also, we received positive feedback from developers, \eg, 
``\emph{This is an outstanding report, thank you so much!}'', 
``\emph{Excellent bug report!}'', 
``\emph{Thanks for the excellent bug report, it was very helpful to fix the bug!}''.

\noindent\textbf{Quality of seed/mutant tests generated by Genie}. We further examined the quality of seed and mutant tests generated by
\tool.
Table~\ref{table:mutation_ability} gives the detailed analysis on the quality of seeds/mutants from \tool. \emph{\#1-O. D. E. Seeds (refined)} and \emph{\#T. P. Seeds} denote the number of seed tests that can generate 1-occurrence distinct error mutants (refined) and true positive mutants, respectively. Thus, \emph{B/A} indicates how many seed tests (in percentage) can generate true positive mutants. We can see, on average, {30.0\% (=42/140)} of seed tests can be successfully mutated to uncover non-crashing bugs.
In Table~\ref{table:mutation_ability}, \emph{T. P. Insert. Pos.} denotes, among those seed tests that can generate true positive mutants, on average how many insertion positions (in percentage) inserted with independent event traces can successfully create true positive mutants. 
We can see the success rate ranges from {18.6$\sim$73.3\%} on these app subjects.
Thus, these results show \tool has reasonable mutation ability to create true positive mutants at both seed test and insertion position levels without any human-provided tests.
We also observe \tool's bug detection ability could be further improved by feeding high-quality seed tests that cover more meaningful app functionalities.
For example, \tool's random seed generation produces only one meaningful seed test for finding the bugs in \emph{SimpleTask}, \emph{Fosdem} and \emph{AnkiDroid}, respectively.

\noindent\textbf{Threats to validity}. Our work may suffer from some threats to validity. First, our technique is only evaluated on 12 apps. To counter this, we select these apps with different features to ensure the generalizability of our evaluation. Table~\ref{table:app_subjects} shows these apps are diverse.
Second, the false positive analysis is manually conducted which may incur imprecision.
To counter this, three co-authors of this paper participate in the analysis and do cross-checking to ensure the correctness. In the future, we will evaluate more apps.

\subsection{Open Challenges}
\label{sec:discussion}

Automatically validating functional correctness of Android apps without explicit oracles
is very challenging.
Our technique is the first to tackle this problem via 
independent view fuzzing.
We face some open challenges.
First, although the independent view property captures a substantial class
of functional bugs, it cannot detect all types of non-crashing functional bugs.
In principle, \tool can only find those non-crashing functional bugs that lead to GUI-level inconsistencies by comparing a seed test and its corresponding mutant tests from the same app version. Thus, \tool cannot detect the non-crashing regression functional bugs (\ie, the app functionality does not work in the current tested app version, but worked in some prior app versions).
One way is to explore other invariant properties of apps or
testing techniques (\eg, using differential testing to detect those regression bugs).	
In addition, \tool cannot detect just-failing functional bugs (\eg, the back-end server is down, so the app cannot function) or
those ad-hoc GUI issues (\eg, color contrast/brightness, text trucked/misalignment issues).

Second, our technique only uses random seed tests for functional validation.
Such seed tests are easy to obtain, but may be unable to cover complicated app functionalities that require human participation.
This constrains the current implementation of \tool to test ``self-contained`` apps (as we discussed in the evaluation setup).
One solution is to leverage
human-provided tests to cover more complicated functionalities.
In the future, we will also explore the possibility of synergistically combining human knowledge with \tool to further improve 
the testing effectiveness and precision.
We believe this is a promising direction to explore.



\section{Related Work}

\noindent{\emph{\textbf{Finding crash bugs of Android apps}}}.
Many \emph{fully automated} GUI testing techniques have been proposed in the literature~\cite{TramontanaAAF19,KongLGLBK19,themis}, including random testing~\cite{monkey,MachiryTN13,appdoctor}, evolutionary algorithms~\cite{MahmoodMM14,mao_sapienz_2016}, symbolic execution~\cite{AnandNHY12,MirzaeiMPEM12,vanderMerwe2012}, model-based testing~\cite{YangPX13,ChoiNS13,AmalfitanoFTTM15,su_stoat_2017,ape_icse2019}, systematic exploration~\cite{a3e,Mirzaei16,ehbdroid,timemachine}, and combinations thereof~\cite{JensenPM13,FanSCMLXP18}.
But all of them can only detect crashes because they use \emph{app exceptions} as the implicit test oracle~\cite{FanSCMLXPS18,why_crash}. They cannot fully automatically uncover non-crashing functional bugs without human-provided test oracles. 

\noindent{\emph{\textbf{Finding non-crashing functional bugs of Android apps}}}.
Finding non-crashing functional bugs depends on oracles.
The current practices heavily rely on manual validation, which incurs substantial cost.
On the other hand, the current research can be
classified into three categories according to the source of oracles.
The first category leverages human-provided oracles to find non-crashing bugs in different strategies.
These oracles are usually encoded in assertions (\eg, \textsc{Thor}~\cite{AdamsenMM15}, \textsc{ChimpCheck}~\cite{LamZC17}, \textsc{AppFlow}~\cite{HuZY18}, ACAT~\cite{RosenfeldKZ18}, 
AppTestMigrator~\cite{BehrangO19}, CraftDroid~\cite{LinJM19}), 
linear-time temporal logic (LTL) formulas (\eg, FARLEAD-Android~\cite{rl_testing}),
or semantic models in Alloy (\eg, \textsc{Augusto}~\cite{MarianiPZ18} which targets web applications).
One special case in this category is the oracles derived from human-written app specifications.
For example, Memon~\etal~\cite{MemonPS00,MemonBN03,XieM07} derives oracles 
(\ie, the expected GUI state) for desktop GUI applications from an initial GUI state of a GUI test based on
human-provided GUI specifications.
\textsc{ReNaLART}~\cite{MalikSKAS21} derives oracles from app specifications written in natural language.
Different from these work, \tool do not require direct or indirect human-provided oracles.

The second category uses differential testing to overcome the oracle problem.
For example, \textsc{SPAG-C}~\cite{LinRCL14} and \textsc{DiffDroid}~\cite{fazzini2017automated}
compare the GUI images or screenshots (explored by the same GUI test) between 
two different mobile devices or platform versions to identify GUI issues.
However, they can only detect device- or platform-specific bugs.

The third category generates automated oracles for a specific class of user interactions.
For example, \textsc{Quantum}~\cite{ZaeemPK14} derives the automated oracles only for app-agnostic interaction features. \textsc{Quantum} assumes an app's GUI page should not change after double-screen-rotations, 
killing-and-restarting or pausing-and-resuming (\textsc{Thor}~\cite{AdamsenMM15} injects similar ``neutral'' events into existing developer tests to do functional testing); or should keep partial GUI elements after zooming-in, zooming out
or scrolling. 
Due to \textsc{Quantum}~\cite{ZaeemPK14} (also \textsc{Thor}~\cite{AdamsenMM15}) can only validate specific app-agnostic features, 
it cannot detect the 34 non-crashing bugs found by \tool because all these bugs are related to app-specific features.
SetDroid~\cite{SunSLDPXS21} can generate automated test oracles for system setting-related, non-crashing defects of Android apps. But it cannot detect the 34 non-crashing bugs found by \tool because all these bugs are not related to system settings.
\textsc{ACETON}~\cite{JabbarvandMM20} uses deep learning to automatically construct test oracles for detecting
energy bugs (which are non-functional bugs), while \tool targets functional bugs.

\noindent{\emph{\textbf{Mutation-based testing for Android apps}}}.
Our approach generates mutant tests by inserting independent event traces, which is quite different from prior mutation-based testing techniques.
Sapienz~\cite{mao_sapienz_2016} shuffles the orders of events between two seed tests to create new tests. EvoDroid~\cite{MahmoodMM14} uses evolutionary algorithms to inject, swap or remove events based on given test cases. Specifically, it 
uses a static UI interface model and program call graphs to do event dependency analysis to ensure the test connectivity.
\textsc{Quantum}~\cite{ZaeemPK14} and \textsc{Thor}~\cite{AdamsenMM15} injects specific neutral events (\eg, double-screen-rotations, pausing-and-resuming) to create new tests. 
Amalfitano \etal~\cite{AmalfitanoRPF18} inject screen rotations 
to stress test Android apps.
TCM~\cite{KorogluS18} uses six different mutation operators (\eg, repeat existing event sequence, remove event delays, change texts) to mutate a seed test.
Other tools, like KREfinder~\cite{Shan16} and APEChecker~\cite{FanSCMLXP18}, injects specific events (\eg, app pausing-and-resuming, quickly start and exit of a GUI page) to help reproduce certain types of errors. In contrast, the mutation strategy in \tool is able to insert generic GUI events via querying the GUI transitional model, which is more flexible and general.

\noindent{\emph{\textbf{Metamorphic Testing}}}.
Our approach leverages the idea of metamorphic testing~\cite{chen1998metamorphic,ChenKLPTTZ18,survey_metamorphic_testing} at the high-level.
The key idea of metamorphic testing is to detect violations of domain-specific metamorphic relations by comparing the outputs between a seed test and its corresponding mutant tests.
It has been widely used in testing many other specification-less software systems, \eg, compilers~\cite{le2014compiler,sun2016finding}, debuggers~\cite{TolksdorfLP19}, and machine learning-based systems~\cite{XieHMKXC11,TianPJR18}.
To the best of our knowledge, we are the first to apply the idea of metamorphic testing to automatically detect non-crashing functional bugs, based on our key observation of independent view property. 
\section{Conclusion}
\label{sec:conclusion}

We have introduced independent view fuzzing, a novel, fully automated approach to detecting 
non-crashing functional bugs of Android apps.
The key idea is to leverage the independent view property 
to manufacture property-preserving mutant tests from a set of seed
tests.
It overcomes the difficulty of functional testing due to lack of test oracles by
checking whether the GUI effects of a seed test are preserved in each of its mutants.
The realization of our idea, \tool, has successfully discovered \totalbugs
non-crashing bugs from twelve real-world apps, 
all of which have been confirmed and 22 were already fixed.
All these bugs were previously unknown and most are long latent.
Our work complements and enhances existing manual testing and fully automated testing for crashes.


\begin{acks}
	We thank the anonymous SPLASH/OOPSLA reviewers for their valuable feedback.
	This work is partially supported by NSFC Project No. 62072178. Zhendong Su and Ting Su were partially supported by a Google Faculty Research Award. Ting Su was also partially supported by a Swiss NSF Spark Award CRSK-2\_190302. Jue Wang was partially supported by NSFC Project No. 61690204.
\end{acks}

\appendix
\section{Appendix}

\subsection{Prevalence of the Indpendent View Property}
\label{sec:appendix_investigation}

\label{sec:bug_investigation}
To validate the prevalence of the independent view property,
we investigated the non-crashing functional bugs of five popular apps, 
\ie, \emph{Ankidroid}~\cite{ankidroid}, 
\emph{AmazeFileManager}~\cite{AmazeFileManager}, 
\emph{AntennaPod}~\cite{AntennaPod},
\emph{k-9mail}~\cite{k9}, \emph{TeamNewPipe}~\cite{NewPipe}.
We focused on all their reported bugs within the recent one-year duration (Sep. 2019--Sep. 2020).
This investigation took the authors more than one-person month. 
We detail the process and the results below.

\noindent\textbf{Apps}.
We selected these five apps since they are popular, diverse and 
well-maintained.
They are released on Google Play and their issue repositories are 
open on GitHub. 
\emph{Ankidroid} is an advanced flashcard learning app 
(3K stars, 5M$\sim$10M installations and first released since Jun. 09);
\emph{AmazeFileManager} is a file manager app 
(3.1K stars, 1M$\sim$5M installations and first released since Jul. 14);
\emph{AntennaPod} is a podcast manager app 
(3.1K stars, 500K$\sim$1M installations and first released since Dec. 11);
\emph{k-9mail} is an email client app
(5.3K stars, 5M$\sim$10M installations and first released since Oct. 08);
and \emph{TeamNewPipe} is a video streaming app
(11.4K stars, 10M$\sim$50M installations and first released since Sep. 15).

\noindent\textbf{Investigation process}.
For each app, we first collected all the bugs 
reported between Sep. 2019 and Sep. 2020 with the issue
labels assigned by developers themselves (\eg, ``bug``, 
``Issue-Bug``).
We chose one-year duration in order to balance the representativeness
of this investigation and the cost of manual analysis.
Table~\ref{table:mini_study} shows the number
of app release versions (denoted by \textbf{\#V}) 
and the number of total bugs (denoted by \textbf{\#T}) 
within this one-year duration.
We then manually identified and excluded those
crash or exception bugs, which are not the focus of this paper.
Next, for the remaining non-crashing bugs, we manually read their
bug reports to check whether they are valid non-crashing functional bugs.
For example, we excluded the non-functional bugs
(\eg, performance or energy issues), usability bugs (\eg,
accessibility issues, color contrast issues, 
misaligned GUI views, truncated texts), feature issues
(problematic feature design) and
compatibility issues (only happened on specific devices).
Last, we tried our best to manually reproduce those 
valid functional bugs by following the reproducing steps
in their bug reports if exist.
For each reproducible functional bug, we manually analyzed
whether it violated the independent view property and
could be detected by the independent view fuzzing approach.

\begin{table}[t]
	\footnotesize
	\newcommand{\tabincell}[2]{\begin{tabular}{@{}#1@{}}#2\end{tabular}}
	\renewcommand{\arraystretch}{1}
	
	\caption{Detailed statistics of investigating the prevalence of independent view property for finding non-crashing functional bugs.}
	\label{table:mini_study}
	\vspace*{-0.5pc}
	\centering
	\begin{adjustbox}{max width=\textwidth}
		\begin{tabular}{|c||c|c|c|c|c|c|c|c|c|}
			\hline
			App Name &\tabincell{c}{\#V} &\tabincell{c}{\#T} & \tabincell{c}{\#N} &\tabincell{c}{\#F}  & \tabincell{c}{\#R} & \tabincell{c}{\#O}  & \tabincell{c}{\#P\\(=\#O/\#F)} \\
			\hline \hline
			\emph{Ankidroid} &25 &215 & 90 & 46  &28 &13 & 28.3\%  \\
			\emph{AmazeFileManager} &10 &70 & 39 & 19  & 14 & 4 & 21.1\% \\
			\emph{AntennaPod} &7 &53 &42 &23 &15 &6 &26.1\% \\
			\emph{k-9mail} &23 &91 &67  &27  &15 &9 &33.3\%  \\
			\emph{TeamNewPipe} &23 &78 &44 &14  &9 &6 &42.8\%  \\
			\hline 
			\textbf{Total} &88  &507  &282 &129  &81 &38 &29.5\% \\
			\hline 
		\end{tabular}
	\end{adjustbox}
\end{table}

\noindent\textbf{Results}.
Table~\ref{table:mini_study} gives the detailed statistics
of this investigation.
\textbf{\#T}, \textbf{\#N}, \textbf{\#F}, \textbf{\#R} and
\textbf{\#O} denote the number of reported bugs in total,
non-crashing bugs, valid functional bugs, reproducible functional
bugs and the functional bugs that are reproducible and
indeed violate the independent view property, respectively.
\textbf{\#P} denotes the percentage of functional bugs that
violate the independent view property (=\textbf{\#O}/\textbf{\#F}).
Note that we have to reproduce a functional bug for confirmation
because the independent view property is GUI-based
(only reading the textual bug reports is not enough for analysis).
Many functional bugs cannot be reproduced because their bug
reports have not provided clear and enough reproducing information
(\eg, steps to reproduce).
We can see, among all the non-crashing functional bugs,
the percentages of violating independent view property
range from 21.1$\sim$42.8\% individually.
Out of 129 non-crashing functional bugs, we identified in these five apps,
38 bugs (29.5\%$\approx$38/129) violated the independent view property. 
Note that these percentages indicate the lower bounds because 
many functional bugs cannot be analyzed (they are not reproducible).
\emph{Thus, it shows validating the independent view property
is applicable to find many non-crashing functional bugs.}

Further, we find these five apps contain developer-written 
regression tests (and oracles), but these tests still miss many 
non-crashing functional bugs (as we can see in Table~\ref{table:mini_study}).
Specifically, \emph{AnkiDroid} has 431 unit and 59 GUI tests;
\emph{AmazeFileManager} has 211 unit and 37 GUI tests;
\emph{AntennaPod} has 197 unit and 55 GUI tests;
\emph{k-9mail} has 585 unit tests; and 
\emph{TeamNewPipe} has 30 unit tests.
\emph{Thus, human-written tests and oracles are usually inadequate to
detect non-crashing functional bugs.}

\bibliographystyle{ACM-Reference-Format}
\bibliography{sample-base}

\end{document}